\documentclass[%
 reprint,
 amsmath,amssymb,
 aps,
]{revtex4-2}

\usepackage{tabularx}
\usepackage{graphicx}
\usepackage{dcolumn}
\usepackage{bm}
\usepackage{csquotes}

\begin{document}

\preprint{APS/123-QED}

\title {Microscopic Gyration with Dissipative Coupling}
\author{Soham Dutta, Arnab Saha}
\affiliation{Department of Physics, University Of Calcutta, 92 Acharya Prafulla Chandra Road, Kolkata-700009, India}

\date{\today}

\begin{abstract}

Microscopic gyrators, including (but not limited to) Brownian gyrators (BGs), require anisotropic fluctuations to perform gyration. It produces a finite current, driving the system out of equilibrium. In a typical BG set-up with an isotropic (i.e. spherical) colloidal particle, the anisotropy sets in by the coupling among space dimensions via an externally applied anisotropic (spherically asymmetric) potential confining the particle and the difference between the temperatures along various space dimensions. The coupling is conservative as it can be derived from a potential. Here, contrary to a typical BG, first we consider an over-damped, anisotropic colloidal particle (a Brownian ellipsoid), trapped in an isotropic (spherically symmetric)  harmonic potential in two dimensions (2D).  The trapping potential being isotropic, cannot couple the space dimensions. Instead, they are coupled by the difference between the longitudinal and transverse frictional drags experienced by the ellipsoid, together with a finite tilt in its orientation due to the chirality of the particle. The coupling can not be derived from a potential. It  is dissipative in nature. It is associated with the intrinsic properties of the particle. We have shown that this dissipative coupling can generate enough anisotropic fluctuations to perform a steady-state gyration in the Brownian scale. Next, going beyond BG, we have considered an inertial, granular, chiral ellipsoid in 2D, subjected to athermal, anisotropic fluctuations. Contrary to the overdamped BG, there is no trapping force confining the granular ellipsoid. However, the coupling between the velocity components of the granular ellipsoid is still dissipative as before. We have shown that being assisted by the dissipative coupling and the anisotropic fluctuations, the inertial, granular ellipsoid can also perform gyration in 2D. Furthermore, we have shown that the dominant contribution towards the gyrating frequency can be attributed to the Coriolis force acting on the granular ellipsoid.  Hence, similar to the Brownian scale, the gyrator in the granular scale is also a tiny autonomous machine that generates a directed motion (gyration) from fluctuations. Although there are fundamental differences between the two.       

\end{abstract}

\maketitle

\section{Introduction}

In the context of simple {\it{thought-vehicles}} with inter-connected motors and sensors which can mimic psychological and behavioral features, Valentino Braitenberg mentioned that: \enquote{The friction, which is nothing but the sum of all the microscopic forces that arise in a situation too messy to be analyzed in detail, may not be quite symmetrical. If the vehicle pushes forward against frictional force, it will deviate from its course. In the long run it will be seen to move in a complicated trajectory...} \cite{braitenberg1986vehicles}. It implies that the inherent asymmetry existing in the friction plays a major role in the directed motion of such vehicles.  How important the asymmetry is will be illustrated in this work. 

Directed motion and non-zero, finite currents are the characteristic features of non-equilibrium systems. In such systems, the current can be generated by a deterministic external \cite{van2004colloids, driscoll2019leveraging,vissers2011lane, samsuzzaman2022reentrant} or motor-like internal drives \cite{wang2015one}. It can also be a fluctuating stochastic force that drives the system out of equilibrium \cite{driscoll2019leveraging, morin2017distortion}. The systems relevant to the current work are the colloidal \cite{everett2007basic,cosgrove2010colloid, dhont1996introduction,ivlev2012complex, evans2019simple} and granular particles \cite{duran2012sands,jaeger1996granular,aranson2006patterns,kadanoff1999built}. They experience asymmetric frictional drags from their surrounding due to the inherent spherically-asymmetric shapes. In the dynamics of colloidal systems, thermal fluctuations are important \cite{schall2006visualizing}. However, in case of granular particles, the effect of thermal fluctuations can be suppressed due to their larger size. Although, the important physics of granular systems can be revealed by externally applied forces, generating athermal fluctuations(e.g. \cite{kudrolli2004size, nowak1998density}). 

The dynamics of a colloidal particle can be represented by stochastic processes. Typically, the dynamics of a single colloidal particle contains inertial forces, viscous forces, and stochastic forces originated from thermal fluctuations and deterministic external fields (if any), such as electromagnetic fields, shear fields, confining potentials, etc. However, the inertial forces are usually dominated by viscous and other forces acting on the colloidal particle. The dissipation of the momentum of a colloidal particle is so fast that inertia can hardly play any role there. Hence, the dynamics of a colloidal particle is predominantly over-damped. In equilibrium colloids, the dissipation caused by the viscous forces is compensated by the stochastic thermal fluctuations. The viscous forces on a colloidal particle is proportional to the velocity of  the particle, where the proportionality constant is the frictional drag co-efficient. The mobility of the particle is the inverse of the drag co-efficient. For a spherical colloidal particle, the frictional drag co-efficient is a scalar, given by the well-known Stokes law \cite{dhont1996introduction,landau2013fluid}. According to the law, the drag co-efficient is proportional to the  viscosity of the surrounding fluid and the size (radius for a spherical particle) of the particle itself. However, in case of an anisotropic  colloidal particle \cite{lee2011recent}, the  frictional drag co-efficient (and hence, the mobility) is a tensor. This simply captures the fact that the friction between the surface of the particle and the surrounding fluid is different along different axes of symmetry of the particle, which is a direct consequence of the inherent shape of the particle. It is this tensorial character of the friction and mobility, for which the motion of the anisotropic colloidal particle along various space dimensions becomes inherently coupled with each other in the (fixed) laboratory frame\cite{dhont1996introduction,han2006brownian,ghosh2020persistence}.        

Contrary to the colloidal particle, the inertia can play a significant role in the dynamics of a granular particle \cite{wylie2003rheology, azema2014internal,vo2020additive,lashgari2016channel,alexander2004effects}. The contribution of the viscous drag due to the surrounding fluid of the granular particle is also significant. However, the inertial and viscous forces can considerably suppress the effect of  thermal fluctuations in the dynamics of the particle\cite{o2004effective,samsuzzaman2022reentrant}. Hence, the physics of granular systems is fundamentally different from that of  the colloidal or any other system which can be equilibrated using a thermal bath via the fluctuation-dissipation relation (FDR) \cite{zwanzig2001nonequilibrium}. Although, one may note here that the interesting physics of granular systems can be revealed by introducing them to athermal stochastic forces externally  [e.g. \cite{briand2018spontaneously}]. A common example is the dynamics of grains on a vibrated plate, where the vibration is due to an athermal noisy force applied externally [e.g. \cite{laroche1989convective, douady1989subharmonic, pak1995effects, wassgren1996vertical}]. In such cases, the dynamics of a granular particle is represented by stochastic processes containing inertial and viscous forces, together with athermal fluctuations. Clearly, in such systems, the strength of the fluctuations can  in general be tuned arbitrarily. It is not necessarily related to the dissipation as in the case of equilibrated colloids. Hence, there are fundamental differences between the stochastic dynamics of a colloidal and a granular particle, primarily due to the nature of the associated fluctuations. However, the tensorial character of the friction (and, mobility), due to the non-spherical geometry of the particles, is qualitatively the same in both cases. The friction tensor can couple the dynamics of the granular particle along the various space dimensions in a similar way as it does in case of a colloidal particle. Here lies the generality of the tensorial aspect of friction, which is a direct and inevitable consequence of the anisotropic shape of the particle, be in the colloidal or in the granular scales.

As previously mentioned, the dynamics of a single colloidal particle and a single granular particle under an externally applied fluctuating force, both can be represented by stochastic processes. The theory of the stochastic processes \cite{gardiner1985handbook} prevails to model the dynamics of the systems like colloids, polymers, granular particles, etc. under various experimental conditions, particularly in the microscopic scales. It remains one of the pivotal topics of research in contemporary soft condensed matter physics. Several works have been conducted regarding single-particle systems, where there exists an innate anisotropy in the  shape of the particle, and as a result, the particle experiences different frictional drags for different modes of its motion (be it translational or, rotational or, both being coupled) [for a review, see \cite{dhont1996introduction}]. When such a spherically asymmetric  system is driven out of equilibrium, the resulting non-equilibrium phenomena becomes explicitly dependent on the shape-asymmetry of the particle via the associated friction tensor.

A prominent example of  non-equilibrium phenomena with a trapped, isotropic (spherical) colloidal particle is the phenomenon of Brownian gyration (BG) \cite{filliger2007brownian}. A typical BG requires anisotropic fluctuation which sets in by the (conservative) coupling of various spatial dimensions via the externally applied anisotropic potential confining the Brownian particle. The particle is driven out of equilibrium by applying different temperatures along different space dimensions and the difference among these temperatures together with the anisotropy of the trap result in the  gyration of the Brownian particle about the minimum of the trapping potential. BG is simultaneously connected to two heat reservoirs of different temperatures. It does not relax to equilibrium but is driven towards a non-equilibrium stationary state (NESS), characterized by a non-zero current.  It has been shown in  \cite{filliger2007brownian} that such a paradigmatic set-up can be mapped to that of a microscopic heat engine consisting of two heat baths, generating work with an emergent torque. The torque vanishes as soon as the temperature difference or the spherical asymmetry of the trap is set to zero. The systematic torque calculated in \cite{filliger2007brownian} is impinged on a physical object residing at the minimum of the trapping potential. Different temperatures along different axis can be established experimentally in several ways \cite{filliger2007brownian}. Recently, BG is experimentally realised in \cite{argun2017experimental}. In case of colloidal micro-heat engines \cite{martinez2017colloidal}, the electrical noise can be used to mimic the thermal noise to set up a heat bath having a different temperature than the fluid in which the colloidal particle resides \cite{martinez2016brownian}. This technique can also be used to have different temperatures along different degrees of freedom of a microscopic gyrator. V. Mancois et. al. obtained the probability density function explicitly and the average angular velocity for both the over-damped and under-damped situations arising in case of a gyrating Brownian sphere, confined in an anisotropic harmonic trap and subjected to two different temperatures along  two space-dimensions \cite{mancois2018two}. The Langevin dynamics \cite{risken1996fokker,van1992stochastic} involved in BG in a confining anisotropic parabolic potential have also been studied in \cite{dotsenko2013two}. The electrical analog of the BG was both experimentally and theoretically realised by \cite{chiang2017electrical}. Here, an autonomous BG was  realised by a simple stochastic electrical circuit with resistors and capacitors. The power spectral densities of such electrical systems have been studied in \cite{cerasoli2022spectral}. In a similar context, the non-equilibrium thermodynamics of such systems (in particular the heat flux and entropy production) have also been studied by \cite{ciliberto2013heat, ciliberto2017experiments}. A gyrating engine in the inertia-less regime with one rotational degree of freedom has been explicitly studied in \cite{siches2022inertialess}, where the emergent torque had an explicit dependence on the angle of rotation. The discussion on the autonomous Stirling engine \cite{izumida2018nonlinear, toyabe2020experimental}  in  \cite{siches2022inertialess} is also contextual. The large-deviation properties and the associated statistics of entropy production in BG have been explicitly explored in \cite{mazzolo2023nonequilibrium}. The works in \cite{dotsenko2022cooperative} have exhibited the co-operative dynamics existing between gyrating molecular tops. The theoretical paradigm for a mesoscopic Brownian heat pump is proposed in \cite{abdoli2022tunable}.  Here, a single charged Brownian particle undergoes gyration on being steered by an external magnetic field. The case where the trapping potential of BG is non-harmonic has been explored in \cite{chang2021autonomous} and while it is time-dependent is studied in \cite{baldassarri2020engineered}. In \cite{squarcini2022fractional}, the BG has been subjected to non-trivial fractional Gaussian noises. The concept of an effective temperature as applicable to BGs has been introduced in \cite{cerasoli2018asymmetry}. The work of \cite{bae2021inertial} shows how inertia and the explicit presence of mass in the dynamical equations can influence the energetics of BG.

Here, we further push the envelop of the research on BG in multiple ways: 

(A) First, in the overdamped regime, instead of considering an anistropic trapping potential, we have considered an isotropic trap that confines an anisotropic, chiral Brownian particle in 2D. Consequently, the dynamical equations of the particle along the space dimensions  are now coupled by the shape-anisotropy of the particle. In particular, they are coupled by the difference of the friction co-efficients along the symmetry axes of the particle. Moreover, due to the chirality of the particle, their angular orientation has a non-zero mean \cite{tsori2020bistable,caprini2023chiral,van2008dynamics,shelke2019transition,kummel2013circular}. This also couples the space dimensions involved in the dynamical equations of the particle. Hence, contrary to the previously mentioned BGs in \cite{filliger2007brownian} and others, the coupling is now dissipative and it depends on the intrinsic properties --- geometry and orientation of the particle. We have shown here that the dissipative coupling, together with the temperature difference along the spatial axes, can produce enough anisotropic fluctuations to generate BG. We emphasize that in the over-damped regime, to exhibit BG with a non-spherical Brownian particle, the anisotropy of the confinement is not necessary. However, the particle is needed to be trapped in an isotropic potential.

(B) Second, we have elevated the concept of BG from position-space to velocity-space by introducing inertia and athermal fluctuations with different strengths along different space dimensions.  This part of the research is particularly relevant for the anisotropic, chiral, granular particles subjected to an external noisy force \cite{arora2021emergent, narayan2007long}. The noisy force has different strengths along different spatial dimensions, however, with the FDR being absent. Nonetheless, we have shown that similar to the Brownian particle, the granular particle can gyrate too and the gyration frequency depends on the difference between the friction coefficients along different symmetry axes of the particle. It also depends on the difference between the strengths of the fluctuations along various space dimensions and on its non-zero mean orientation (due to chirality). It turns out that there are fundamental differences between the velocity-space gyration (granular case) and position-space gyration (colloidal / Brownian case). In case of the granular ellipsoid, the particle does not even require any external trap to gyrate, whereas in the colloidal ellipsoid case, a spherically symmetric trap is needed.
Furthermore, we have shown that in case of the granular particle, the Coriolis force (arising due to the rotating body frame) \cite{coriolis1835memoire, jose2000classical} contributes significantly to the gyration. This invokes the possibility of the emergence of Coriolis force from the un-biased fluctuating forces in a table-top set-up of a granular system, under random, vertical vibrations\cite{arora2021emergent,blair2003vortices,yamada2003coherent,aranson2006patterns,narayan2007long}.

The paper is organized as follows. We first consider the overdamped Brownian motion of an anisotropic colloidal particle in the form of an ellipsoid, trapped in an isotropic harmonic potential in 2D. The average angular orientation of the ellipsoid is non-zero. We will calculate the complete probability distribution, mean gyration current and frequency analytically for a Brownian ellipsoid in 2D, which is slightly spherically asymmetric and a little tilted from its long axis.
In the second halve of the paper, we have dealt with the underdamped case of a granular ellipsoid in 2D, again with a finite mean orientation. It is subjected to externally applied athermal fluctutations with different strengths along different space dimensions. Gyration is observed for this case as well. The complete probability distribution, mean gyration current and frequency of gyration are calculated explicitly, in the small coupling limit. Small coupling implies a non-zero but small average angle of orientation and a slight spherical asymmetry of the particle. A fair agreement is observed between the theoretically calculated non-equilibrium quantities and their numerical counterparts in both the cases.

\section{Gyration Of A Brownian Ellipsoid in 2D}

\subsection{The Model}
\label{model}

\begin{figure}[htp]
    \centering
    \includegraphics[width=5cm]{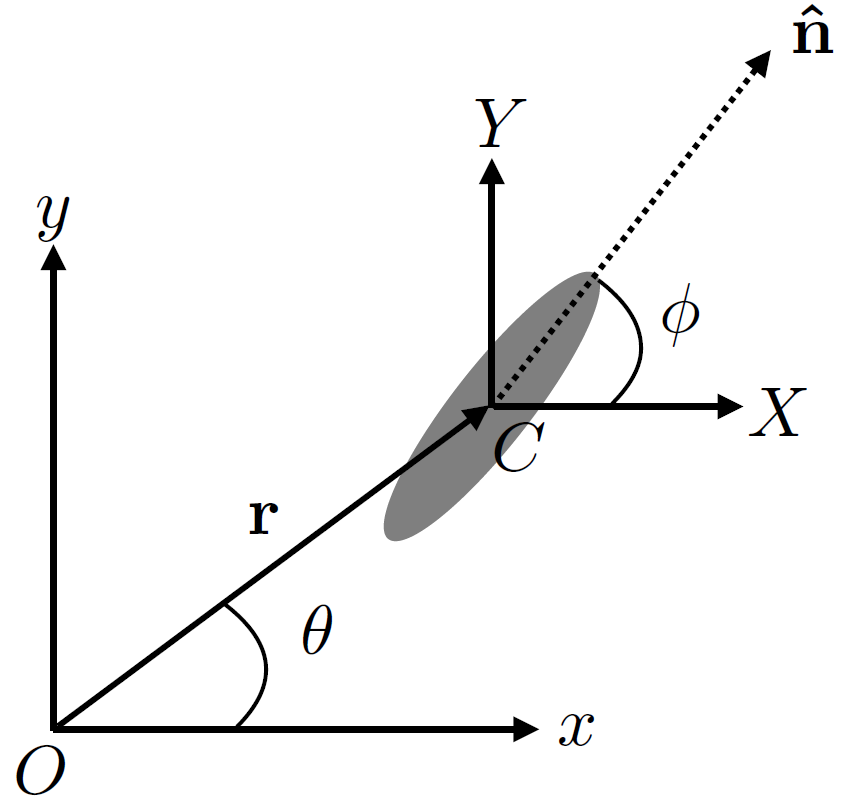}
    \caption{The Body Frame is denoted by  $(X,Y)$ and the Lab Frame is denoted by $(x,y)$. The origin of the body frame is the centre-of-mass (C) of the ellipsoid. Here, $(r, \theta)$ are the polar co-ordinates of  C with respect to the lab frame. Equivalently, in the Cartesian system, the coordinates of C in the lab frame can be denoted by $(x,y)$. The angle $\phi$ denotes the angular orientation of the ellipsoid. It is the angle subtended by the unit vector $\hat {\bf n}$ (along the major axis of the ellipsoid) on the body axis $X$ at C.} 
\label{ellipsoid}    
\end{figure}

We consider a 2D system consisting of a colloidal ellipsoid of unit mass, suspended in a highly viscous fluid where the rate of change of momentum of the particle is  negligibly small. Since the dynamics of the particle is restricted to the $xy$-plane (see Fig.[\ref{ellipsoid}]), its centre of mass (denoted by C in Fig[\ref{ellipsoid}]) is moving across the plane only. Furthermore, the particle is free to rotate around C with the axis being perpendicular to the $xy$-plane and passing through C. This rotation is denoted by the angle $\phi$ in Fig[\ref{ellipsoid}]. It determines the angular orientation of the particle. The overdamped equation of motion of the particle will involve only the coordinates of C at time $t$, denoted by $(x,y)$ and the angle of orientation $\phi$. In Fig[\ref{ellipsoid}], the coordinates $(x,y,\phi)$, that completely specify the overdamped dynamics of the ellipsoid in 2D, are depicted. The time evolution of the dynamical variables with respect to the lab-frame is given by the overdamped Langevin equations as (\cite{han2006brownian} \cite{ghosh2020persistence}):

 \begin{eqnarray}
&&\partial_t x_i=-\sum\limits_{j}\Gamma_{ij}(\phi) \frac{\partial U}{\partial x_j}+f_i(t)\\
&&\partial_t \phi=\Gamma_r\mathcal{M}
\label{eomm}
\end{eqnarray}
where, $i\in (1,2)$, $j\in (1,2)$, $x_1\equiv x$, and, $x_2\equiv y$. 

According to Eq.[\ref{eomm}], the particle is subjected to three distinct forces during its translation---

(i) First, the particle is trapped in a spherically symmetric 2D harmonic potential, $U(x,y)=\frac{k}{2}(x^2+y^2)$. Being spherically symmetric, it cannot couple $x$ and $y$ in the equation of motion.

\begin{figure}[htp]
    \centering
    \includegraphics[width=5cm]{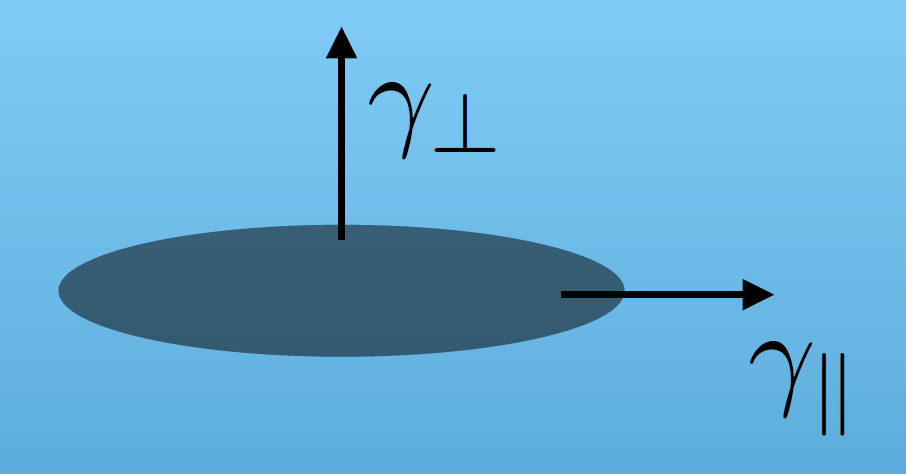}
    \caption{Longitudinal ($\gamma_\parallel$) and Transverse ($\gamma_\perp$) Friction Co-efficients of an Ellipsoidal Particle}
    \label{EllipseFriction}
\end{figure}

(ii) Second, the viscous drag  $-\gamma_{ij}v_j$, where $\gamma_{ij}$ is the friction tensor where $v_j=\partial_tx_j$. In 2D, $\gamma_{ij}$ is a $2\times 2$ matrix, the inverse of which is the mobility tensor represented by $\Gamma_{ij}$. The components of $\Gamma_{ij}$ 
depend on the geometry and the orientation angle $\phi$ of the particle as \cite{dhont1996introduction},

\begin{eqnarray}
\nonumber
\Gamma_{ij}(\phi)&=&\Gamma_{\parallel}(\hat{\bf{n}} \otimes \hat{\bf{n}})+\Gamma_{\perp}[\delta_{ij}-(\hat{\bf{n}} \otimes \hat{\bf{n}})]\\ 
 &=&\Gamma\delta_{ij}+\frac{\Delta\Gamma}{2}M_{ij}(\phi)
\label{mobility}
\end{eqnarray}
Here, $\Gamma_{\parallel, \perp}=\frac{1}{\gamma_{\parallel,\perp}}$, where $\gamma_{\parallel}$ and $\gamma_{\perp}$ are the frictional drag co-efficients of the ellipsoidal particle along the directions parallel and perpendicular to its major axis respectively (see Fig.[\ref{EllipseFriction}]). Here, $\gamma_{\perp}>\gamma_{\parallel}$, and they depend on the ratio of the lengths of major and minor axis of the ellipsoid particle \cite{dhont1996introduction}. Here, $\Gamma \equiv \frac{1}{2}\left(\Gamma_{\parallel}+\Gamma_{\perp}\right)$,  $\Delta\Gamma\equiv\Gamma_{\parallel}-\Gamma_{\perp}$, and $\hat{\bf{n}}\equiv \begin{bmatrix} \cos\phi\\\sin\phi \end{bmatrix}$  is the unit vector (in 2D) along the major axis of the ellipsoidal particle, and $\hat{\bf{n}} \otimes \hat{\bf{n}}=n_in_j=\begin{bmatrix} \cos^2\phi & \cos\phi\sin\phi\\\cos\phi\sin\phi & \sin^2\phi \end{bmatrix}$. Here, $M_{ij}(\phi)=\begin{bmatrix} \cos2\phi & \sin2\phi\\\sin2\phi & -\cos2\phi \end{bmatrix}$. Explicitly, the mobility tensor then takes the form of a $2\times2$ matrix with $\phi$-dependent elements given by:

\begin{eqnarray}
\nonumber
\Gamma_{xx}=\Gamma_{\parallel}\cos^2\phi+\Gamma_{\perp}\sin^2\phi\\
\nonumber
\Gamma_{yy}=\Gamma_{\parallel}\sin^2\phi+\Gamma_{\perp}\cos^2\phi\\
\Gamma_{xy}=\Gamma_{yx}=\Delta\Gamma \sin\phi\cos\phi
\end{eqnarray} 

Clearly, if the particle is spherically symmetric, then $\Delta\Gamma=0$ and $\Gamma_{ij}$ becomes diagonal with $\Gamma_{\perp}=\Gamma_{\parallel}=\Gamma$. Hence, $\Delta\Gamma$ is a measure of the spherical asymmetry of the particle.  Later in this paper, it will become evident that the off-diagonal terms (containing $\Delta \Gamma$) couples $x$ and $y$ in the equation of motion, which is essential for the gyration to occur. The coupled equations for the translational motion of the Brownian ellipsoid in 2D can now be explicitly given as, 

\begin{eqnarray}
\nonumber
\dot{x}&=&-\Gamma_{xx}\frac{\partial U}{\partial x}-\Gamma_{xy}\frac{\partial U}{\partial y}+f_x(t)\\
\dot{y}&=&-\Gamma_{yx}\frac{\partial U}{\partial x}-\Gamma_{yy}\frac{\partial U}{\partial y}+f_y(t)
\label{eom2}
\end{eqnarray}

One may note here that the Langevin equations for $x$ and $y$ are coupled with each other via the matrix element $\Gamma_{xy}(\phi,\Gamma_{\parallel},\Gamma_{\perp})$, as mentioned earlier. However, the Langevin equation for $\phi$ has no dependence on its translational counterparts, $x$ and $y$.
 
(iii) Fluctuating forces and torques : The noises, $f_x(t)$ and $f_y(t)$, are the zero-mean Gaussian, white random force components along $x$ and $y$ respectively. The temperatures along $x$ and $y$ are $T_x$ and $T_y$ respectively with $T_x\neq T_y$. The strengths of the noises maintain the fluctuation-dissipation relation (FDR) with the respective temperatures. Hence, in the units of $k_B$,

\begin{eqnarray}
\nonumber
\langle f_x(t) \rangle&=&\langle f_y(t) \rangle=0\\
\nonumber
\langle f_x(t) f_x(t')\rangle &=&2T_x\Gamma_{xx}\delta(t-t')\\
\langle f_y(t) f_y(t')\rangle&=&2T_y\Gamma_{yy}\delta(t-t')
\label{corr}
\end{eqnarray}

Similarly, in the dynamics of $\phi$ given by Eq.[\ref{eomm}], the mobility $\Gamma_r$ is introduced due to the viscosity of the surrounding fluid in which the particle allows itself to rotate. As there is only one orientational degree of freedom $\phi$, unlike $\Gamma_{ij}$, $\Gamma_r$ is a scalar. Note that for the coupling between $x$ and $y$ to survive statistically, we would also need $\langle\sin\phi\cos\phi\rangle=\int d\phi \sin\phi\cos\phi P(\phi)\neq 0$, where $P(\phi)$ is the distribution of $\phi$. Clearly, the condition $P(\phi)\neq P(-\phi)$ is necessary to achieve this. Without employing an external torque on the particle, this can be acquired if the particle is intrinsically chiral.

In the dynamical equation of $\phi$ given by Eq.[\ref{eomm}], $\mathcal{M}$ is the total torque involved in the rotational dynamics of the particle. As mentioned earlier, $\mathcal{M}$ is fluctuating in time such that $\langle \sin2\phi\rangle \neq 0$. It implies that the orientation of the particle is biased and the particle is tilted at a finite angle with respect to $\phi=0$. It is possible when the particle is inherently chiral, due to which $\mathcal{M}$ becomes a sum of a deterministic torque generated by the internal chiral degrees of freedom of the particle and the thermally generated stochastic torque, $f_r=\sqrt{2\gamma_rk_BT_r}\xi_r(t)$, where $\gamma_r=\Gamma_r^{-1}$, $T_r$ is the temperature associated with the dynamics of $\phi$ ($T_r$ can be, in general, different from $T_x$ and $T_y$) and $\xi_r$ is a Gaussian, white random variable such that $\langle \xi_r(t)\rangle = 0$, and $\langle\xi_r(t)\xi_r(t')\rangle=\delta(t-t')$. Here, we would like to emphasize that chirality is an inherent internal property of the particle, just as $\Delta\Gamma$ is. The coupling $\Gamma_{xy}=\Gamma_{yx}$ between the dynamics of $x$ and $y$ depends only on the characteristic features of the particle and not on the trap in which it is confined.

\subsubsection{Small fluctuation in $\phi$}

In general, $\mathcal{M}$ determines $P(\phi)$. If  the particle is isotropically oriented, i.e. $\mathcal{M}=\sqrt{2\gamma_rk_BT_r}\xi_r(t)$ only,  $P(\phi)$ will also be a zero-mean Gaussian noise as $\xi_r(t)$, leading towards $\langle \phi\rangle=0$. However, as mentioned before, here we consider the particle to be slightly chiral or intrinsically anisotropic in its orientation. Therefore, apart from the random torque, $\mathcal{M}$ will have another small contribution (say `$\Lambda$'), such that $\mathcal{M}=\Lambda+\sqrt{2\gamma_rk_BT_r}\xi_r(t)$. Thus, $P(\phi)$ becomes slightly asymmetric about $\phi=0$, which leads to $\langle\phi\rangle \equiv \delta\phi\neq 0$. Here, $\delta\phi$ is so small that except  the terms $\mathcal{O}(\delta\phi)$, all terms $\mathcal{O}(\delta\phi^2)$ and those of higher order in Eq.[\ref{eom2}] are negligibly small. As a result,  after taking the average over $\phi$, Eq.[\ref{eom2}] becomes,

\begin{eqnarray}
\nonumber
\dot{x}=-\frac{k}{\gamma_\parallel} (x+c_1 y)+\sqrt{\frac{2T_x}{\gamma_\parallel}}\xi_x(t)\\
\dot{y}=-\frac{k}{\gamma_\perp} (y+c_2 x)+\sqrt{\frac{2T_y}{\gamma_\perp}}\xi_y(t)
\label{eom3}
\end{eqnarray}

where, $c_1=\frac{\delta\phi\Delta\Gamma}{\Gamma_{\parallel}}$ and $c_2=\frac{\delta\phi\Delta\Gamma}{\Gamma_{\perp}}$ are the dimensionless coupling constants in this limiting case. Here we consider, $\langle\xi_l\rangle=0$, and, $\langle\xi_l(t)\xi_l(t')\rangle=\delta(t-t')$, where $l\in(x,y)$ to maintain Eq.[\ref{corr}]. 

We emphasize here that the coupling between $x$ and $y$ survives in Eq.[\ref{eom3}] through the geometric anisotropy $(\Delta\Gamma)$ of the Brownian ellipsoid and the associated mean orientation $(\delta\phi)$, both being the intrinsic features of the particle. As introduced before, the coupling is a dissipative one, due to the explicit presence of the friction in the coupling constants. One may also note here that as the above equations are linear, the distribution of $x(t)$ and $y(t)$ will be Gaussian due to $\xi_x(t)$ and $\xi_y(t)$.

\subsubsection{Small $\Delta\Gamma$ limit}

One can simplify Eq.[\ref{eom3}] even further by taking $\Delta\Gamma\rightarrow 0$ but not exactly equal to zero. In this limit, $\gamma_{\perp}$ and $\gamma_{\parallel}$ are so close to each other that both can be replaced by their average, $\gamma=\frac{\gamma_{\perp}+\gamma_{\parallel}}{2}$. Thus, we get

\begin{eqnarray}
\nonumber
\dot{x}=-\frac{k}{\gamma} (x+\epsilon y)+\sqrt{\frac{2T_x}{\gamma}}\xi_x(t)\\
\dot{y}=-\frac{k}{\gamma} (y+\epsilon x)+\sqrt{\frac{2T_y}{\gamma}}\xi_y(t)
\label{eom4}
\end{eqnarray}

Here,
\begin{eqnarray}
 \epsilon=\frac{\delta\phi\Delta\Gamma}{\Gamma}   
\end{eqnarray}

where, $\Gamma=\frac{\Gamma_{\perp}+\Gamma_{\parallel}}{2}$ is the mean mobility. This simplification reduces the number of parameters of the system and helps us to understand the essential physics better. Moreover, it ensures $|\epsilon|< 1$, a condition required for the particle to be bounded within the trap. 

\subsection{Results}

In the following few sections, we will show that the Brownian ellipsoid can exhibit gyration while elucidating the method step-by-step to obtain an expression for the average angular velocity $(\omega)$ of the gyration. For simplicity, we will consider Eq.[\ref{eom4}] as the $\phi-$averaged
equation of motion of the centre-of-mass of the ellipsoid, while calculating $\omega$.

\subsubsection{Covariance Matrix and Position Variances}

The dynamics of the position probability distribution of the centre-of-mass of the Brownian ellipsoid, $P(x,y,t)$ is given by the Smoluchowski equation \cite{van1992stochastic, chandrasekhar1943stochastic, risken1996fokker, gardiner1985handbook},  

\begin{eqnarray}
\frac{\partial P(x,y,t)}{\partial t}=-\bf{\nabla.J}
\label{probdyn}
\end{eqnarray}

where, the probability current density ${\bf{J}}$ is given by,

\begin{eqnarray}
J_i=\sum\limits_{j}\left(A_{ij}x_j P-\frac{1}{2}B_{ij}\frac{\partial P}{\partial x_j}\right)
\label{current}
\end{eqnarray}

Here, $A_{ij}$ and $B_{ij}$ are the co-efficient matrices given by,

\begin{eqnarray}
A &=&-\frac{k}{\gamma}\begin{bmatrix} 1 & \epsilon \\ \epsilon & 1 \end{bmatrix}\\
B &=&-\frac{2}{\gamma}\begin{bmatrix} T_x & 0 \\ 0 & T_y \end{bmatrix}
\end{eqnarray}

From Eq.[\ref{probdyn}], the dynamics of the co-variance matrix, $\Sigma=\begin{bmatrix} \sigma_x & \sigma_{xy} \\ \sigma_{yx} & \sigma_y\end{bmatrix}$, where $\sigma_x=\langle x^2 \rangle$, $\sigma_y=\langle y^2 \rangle$, $\sigma_{xy}=\langle xy \rangle$, and $\sigma_{yx}=\langle yx \rangle$, can be written as \cite{van1992stochastic}

\begin{eqnarray}
    \frac{d\Sigma}{dt}=A\Sigma+\Sigma A^T+B
    \label{covdyn}
\end{eqnarray}

Here, $A^T$ denotes transpose of $A$. As there is no bias in the motion of the particle along any of the two independent directions of translation ($x$,$y$), we can set $\langle x\rangle=0=\langle y\rangle$. For stationary condition, $\frac{d\Sigma}{dt}=0$, which leads to the steady state solutions for the moments as functions of system parameters $T_x,T_y,k,\epsilon$,

\begin{eqnarray}
\nonumber
&&\sigma_x(T_x,T_y,k,\epsilon)=\frac{2T_x+\epsilon^2(T_y-T_x)}{2k(1-\epsilon^2)}\\
\nonumber
&&\sigma_y(T_x,T_y,k,\epsilon)=\frac{2T_y+\epsilon^2(T_x-T_y)}{2k(1-\epsilon^2)}\\
\nonumber
&&\sigma_{xy}(T_x,T_y,k,\epsilon)=\sigma_{yx}(T_x,T_y,k,\epsilon)=
\\ &&
-\frac{\epsilon(T_x+T_y)}{2k(1-\epsilon^2)}
\label{moments}
\end{eqnarray}

where, the terms higher than  $\mathcal{O}(\epsilon^2)$ are neglected. This readily shows that the covariance matrix $\Sigma$ is symmetric. It should be noted here that to obtain finite moments,  $|\epsilon|$ should be less than unity which is ensured by taking $\Delta \Gamma$  and $\delta\phi$ both quite small. One may also note here that $\sigma_{xy}$ is non-zero only when the coupling parameter $\epsilon$ is non-zero. 

The ratio of $\sigma_x$ and $\sigma_y$  is a characteristic quantity, which can serve as a measure of anisotropy in the system under consideration \cite{dotsenko2013two}. From Eq.[\ref{moments}], this ratio is given by,

\begin{equation}
\mathcal{R}(T_x,T_y,\epsilon)\equiv\frac{\sigma_x}{\sigma_y}=\frac{2-\epsilon^2(1-\frac{T_y}{T_x})}{2(\frac{T_y}{T_x})+\epsilon^2(1-\frac{T_y}{T_x})}
\label{ratio}
\end{equation}

\begin{figure}[htp]
    \centering
    \includegraphics[width=8cm]{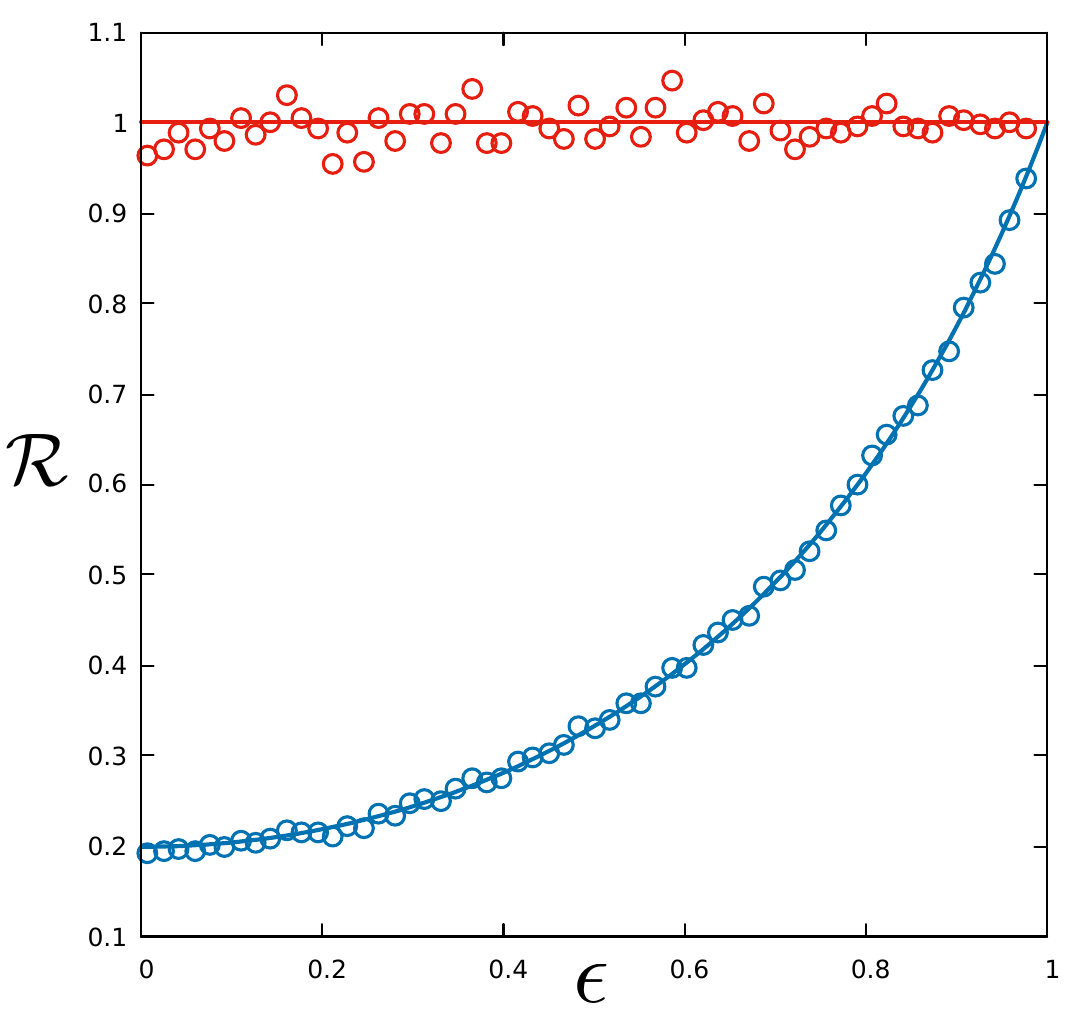}
    \caption{Variation of $\mathcal{R}$ with $\epsilon$. The red and blue circles are the numerical data obtained for $T_y=1$ and $T_y=5$ respectively (with $T_x=1$ for both the cases). The solid lines denote the analytical expressions.}
    \label{moments2}
\end{figure}

In Fig.[\ref{moments2}], $\mathcal{R}$ is plotted with $\epsilon$ ranging from $0$ to $1$. From the equations of motion in Eq.[\ref{eom4}] and the expressions of the moments in Eq.[\ref{moments}], it is clear that as $\epsilon=0$, the dynamics along $x$ and $y$ becomes decoupled, $\sigma_{xy}=0$ and $\mathcal{R}=\frac{T_x}{Ty}$. We confirm this by simulating the system (see Fig.[\ref{moments2}]).  The expressions of the moments in Eq.[\ref{moments}] is apparently blowing up as $\epsilon\rightarrow 1$. Note that the calculation of the moments is correct only when $\epsilon$ is much smaller than $1$. However, the ratio of the moments  as defined in Eq.[\ref{moments}] remains finite even in the limit where $\epsilon\rightarrow 1$. In particular, $\lim_{\epsilon\rightarrow 1}\mathcal{R}=1$. To confirm this, we simulate the system and calculate the ratio numerically for varying $\epsilon$  between $0$ to $1$. This matches well with the theoretical results (see Fig.[\ref{moments2}]).

\subsubsection{Stationary Probability Distribution Function, $P(x,y)$}

As the two equations in Eq.[\ref{eom4}] are linear, the stationary probability distribution function for position, which is also the solution of Eq.[\ref{probdyn}], will be a normalized multi-variate Gaussian distribution, given by:

\begin{equation}
P({\bf{z}})=\frac{e^{-\frac{1}{2}[({\bf{z}}-\langle{\bf{z}}\rangle)^T \Sigma^{-1}({\bf{z}}-\langle{\bf{z}}\rangle)]}}{2\pi\sqrt{{\text{Det}}\Sigma}}\equiv Ae^{-p(x,y)} 
\end{equation}

where, $\bf{z}=\begin{bmatrix} x\\y \end{bmatrix}$, the normalisation constant is denoted by $A$ and the argument of the exponential is denoted by $p(x,y)$. The determinant and inverse of the covariance matrix $\Sigma$ can be calculated as:

\begin{eqnarray}
&&\text{Det}\Sigma=\frac{1}{4k^2(1-\epsilon^2)}[4T_xT_y+(T_y-T_x)^2\epsilon^2]\\
&&\Sigma^{-1}=\frac{2k}{4T_xT_y+\epsilon^2(T_y-T_x)^2}\times \\ \nonumber &&
\begin{bmatrix} 2T_y-(T_y-T_x)\epsilon^2 & (T_x+T_y)\epsilon \\ (T_x+T_y)\epsilon & 2T_x+(T_y-T_x)\epsilon^2\end{bmatrix}
\label{detinv}
\end{eqnarray}

where, the terms of order higher than $\mathcal{O}(\epsilon^2)$ are neglected. Therefore, the normalization constant can be obtained as,

\begin{equation}
A\equiv\frac{1}{2\pi\sqrt{{\text{Det}}\Sigma}}=\frac{k}{\pi}\sqrt{\frac{1-\epsilon^2}{4T_xT_y+\epsilon^2(T_y-T_x)^2}}
\end{equation}

and,

\begin{equation}
p(x,y)=\frac{1}{2}[({\bf{z}})^T \Sigma^{-1}({\bf{z}})]=\Delta_1 (x^2) + \Delta_2 (y^2) +\Delta_3 (xy)
\end{equation}

where,

\begin{eqnarray}
\nonumber
\Delta_1=k \frac{2T_y+\epsilon^2(T_x-T_y)}{4T_xT_y+\epsilon^2(T_x-T_y)^2}\\
\nonumber
\Delta_2=k \frac{2T_x+\epsilon^2(T_y-T_x)}{4T_xT_y+\epsilon^2(T_x-T_y)^2}\\
\Delta_3=k \frac{2\epsilon(T_x+T_y)}{4T_xT_y+\epsilon^2(T_x-T_y)^2}
\end{eqnarray}

Note that for $\epsilon=0$ and $T_x=T_y=T$ (that is, at equilibrium), $P(x,y)$ reduces to the usual Maxwell-Boltzmann distribution. If $\epsilon=0$ but $T_x\neq T_y$ then $x$ and $y$ are decoupled. Different degrees of freedom of the system are equilibrated with different thermal baths at different temperatures and as they are not coupled with each other, there is no heat flux between the baths via the system. Therefore, in this case, $P(x,y)=P^{\text{eq}}(x,T_x,k)P^{\text{eq}}(y,T_y,k)$.  

\begin{figure}[htp]
    \centering
    \includegraphics[width=8cm]{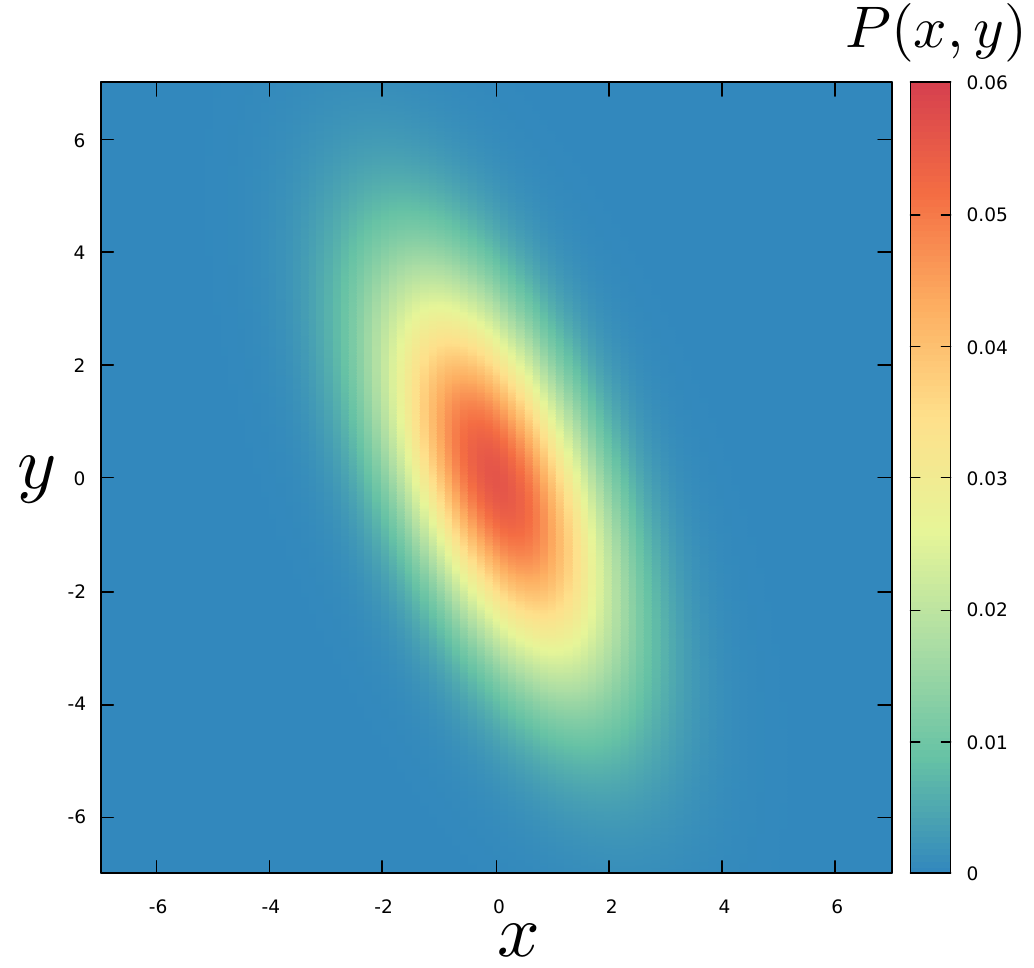}
    \caption{Contour plot of $P(x,y)$, for $k=1, \epsilon=0.5, T_x=1, T_y=5$. Here, the colour scale denotes the various values of $P(x,y)$.}
\end{figure}

When both $\epsilon\neq 0$ and $T_x\neq T_y$, gyration occurs. Calculating the average angular velocity $(\omega)$ for the gyration will be simpler if $P$ can be represented in polar coordinates. In polar co-ordinates, $p(x,y)$ is transformed as:

\begin{equation}
q(r,\theta)=r^2[\Delta_+(1+\epsilon \sin2\theta)+\Delta_- \cos2\theta]
\end{equation}

where,
\begin{eqnarray}
\Delta_+ = \frac{\Delta_1+\Delta_2}{2}\\
\Delta_- = \frac{\Delta_1-\Delta_2}{2}
\end{eqnarray}

Hence, by exponentiating $q$, we obtain the probability density for position of the center-of-mass of the ellipsoid in polar coordinates, $P(r,\theta)$, which will be used to evaluate the current densities in the next section.  

\subsubsection{Stationary Probability Current Density}

In the polar co-ordinate system, using Eq.[\ref{current}], the components of the probability current density can be re-written as,

\begin{eqnarray}
\nonumber
J_x(r,\theta)&=&-\frac{k r}{\gamma}(\cos\theta+\epsilon \sin\theta)P\\
&-&\frac{T_x}{\gamma}\left(\cos\theta\frac{\partial P}{\partial r}-\frac{\sin\theta}{r}\frac{\partial P}{\partial \theta}\right)
\end{eqnarray}
\begin{eqnarray}
\nonumber
J_y(r,\theta)&=&-\frac{k r}{\gamma}(\sin\theta+\epsilon \cos\theta)P\\
&-&\frac{T_y}{\gamma}\left(\sin\theta\frac{\partial P}{\partial r}+\frac{\cos\theta}{r}\frac{\partial P}{\partial \theta}\right)
\end{eqnarray}

where, $P=P(r,\theta)$. The total probability current density given by, ${\bf J}=J_r(r,\theta)\boldsymbol{\hat{r}}+J_\theta(r,\theta)\boldsymbol{\hat{\theta}}$, consists of two components : a radial and a tangential component. The tangential component is given by, $J_\theta(r,\theta)=J_y(r,\theta)\cos\theta-J_x(r,\theta)\sin\theta$. After evaluating all the partial derivatives of $P(r,\theta)$, the explicit form of the tangential  current density can be obtained as:

\begin{eqnarray}
\nonumber
J_{\theta} (r,\theta)&=& \frac{r\epsilon(T_y-T_x)}{\gamma}\times(\Delta_+(1+\epsilon \sin2\theta)+\Delta_- \cos2\theta)\\
&&\times P(r,\theta)
\label{current2}
\end{eqnarray}

Clearly, as long as $T_x\neq T_y$ and $\epsilon>0$, $J_{\theta}$ is non-zero and the system is in a non-equilibrium steady state.

\subsubsection{Average Angular Velocity : Frequency of Gyration}

The angular velocity $\boldsymbol{\omega}$ for a given value of the particle's position, located at ${\bf{r}}$ on a plane, and moving with a linear velocity ${\bf{v}}$ along this plane, can be written as a vectorial cross-product in the form: $\boldsymbol{\omega}=\frac{\bf{r}\times\bf{v}}{r^2}$, directed along the normal to the plane. The ensemble average of the magnitude of $\boldsymbol{\omega}$ will then be given as:

\begin{equation}
\omega =|\int d^2{\bf r}\left(\frac{{\bf{r}}\times{\bf{v}}}{r^2}\right)P(r,\theta)|=\int_0^{2\pi} d\theta \int_0^{\infty}dr J_\theta(r,\theta)
\end{equation}

The average angular velocity (or, in other words, frequency of gyration) can now be explicitly evaluated by plugging in the tangential current density from Eq.[\ref{current2}] as:

\begin{equation}
\omega=\frac{\epsilon k (T_y-T_x)}{\gamma}\sqrt{\frac{1-\epsilon^2}{4T_xT_y+\epsilon^2 (T_y-T_x)^2}}
\label{omega1}
\end{equation}

If one neglects the terms $\mathcal{O}(\epsilon^2)$ in the above equation, the average angular velocity takes the form, $\omega=\frac{\epsilon k}{2\gamma}(\sqrt{\frac{T_y}{T_x}}-\sqrt{\frac{T_x}{T_y}})$.  

\begin{figure}[htp]
    \centering
    \includegraphics[width=8cm]{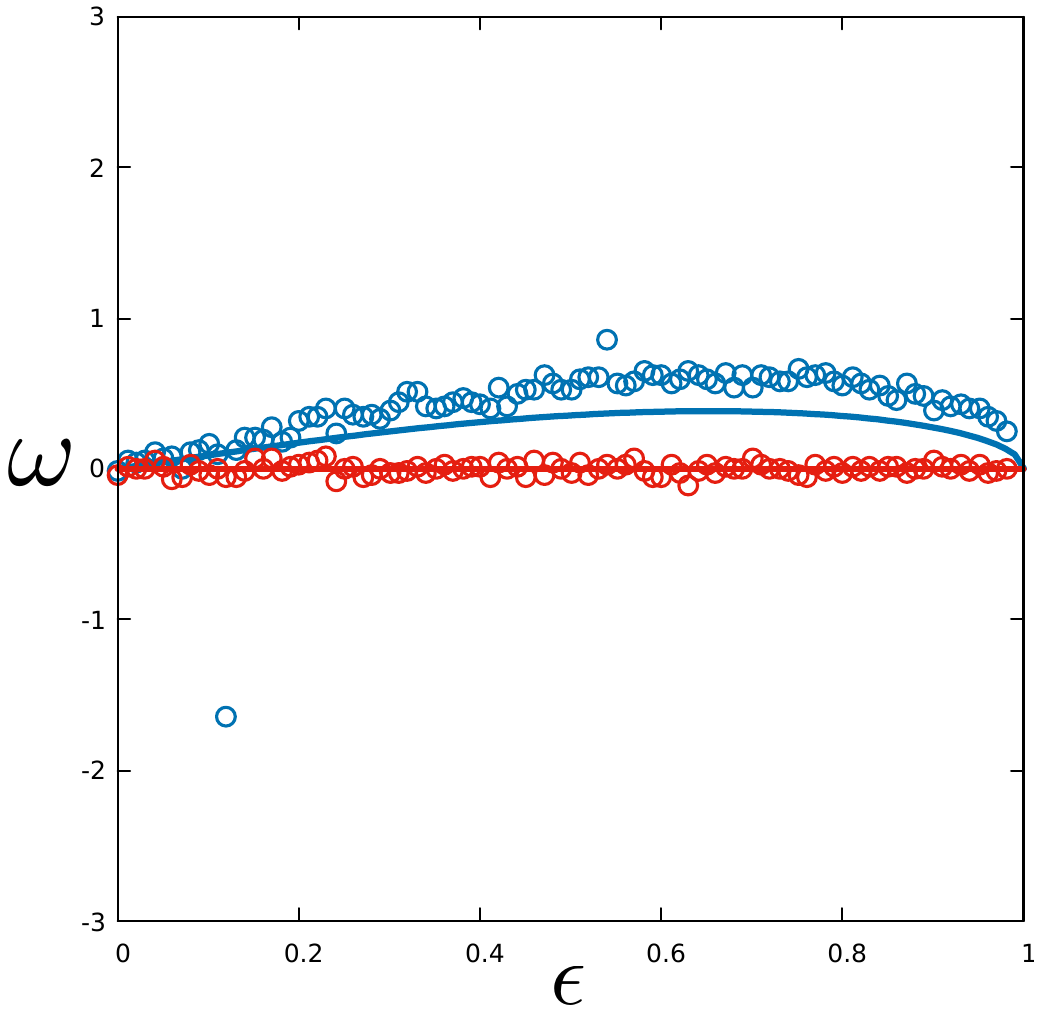}
    \caption{Variation of $\omega$ with $\epsilon$. The red and blue circles are the numerical data obtained for $T_y=1$ and $T_y=5$ respectively (with $T_x=1$, $k=1$, $\gamma=1$ for both the cases). The solid lines denote the analytical expressions.}
    \label{OverdampedOmega}
\end{figure}

First, one may note in the expression that if $T_x=T_y$, $\omega=0$ for any $\epsilon$ between $0$ and $1$. As the theoretical calculation is developed for $|\epsilon| < 1$ such that one can neglect terms higher than $\mathcal{O}(\epsilon^2)$, we need to confirm this by simulating the system. In Fig.[\ref{OverdampedOmega}], we have plotted  $\omega$  from both theory and simulation for $T_x=T_y$ and they match well. This result is intuitive because when $T_x=T_y$, there is no heat flux that can drive the system out of equilibrium, and hence, the particle is unable to gyrate.  

Second, for $T_x \neq T_y$, the  system is in contact with two heat sources of different temperatures. The system is not relaxing to an equilibrium, but is driven towards a non-equilibrium stationary state. Hence, the particle can undergo gyration. The gyration current ${\bf{J}}$ and the frequency $\omega$  become non-zero.  Eq.[\ref{omega1}] clearly shows that when the bath temperatures are not equal, $\omega\neq 0$ for all $\epsilon$ between $0$ to $1$, except at $\epsilon=0$ and $\epsilon=1$. Also, the sense of rotation (clockwise or anti-clockwise) gets flipped when $T_x$ becomes greater than $T_y$.  

At $\epsilon=0$, even when $T_x\neq T_y$, $\omega=0$ according to Eq.[\ref{omega1}]. Hence, the gyration is absent in such a scenario and the system remains at equilibrium. This is because when $\epsilon=0$, $x$ and $y$ degrees of freedom of the system become decoupled and there is no exchange of heat between them, which can drive the system out of equilibrium. Therefore, each of the degrees of freedom representing the system is individually equilibrated at different temperatures. For very small $\epsilon$, when only $\mathcal{O}(\epsilon)$ term is considered, theory predicts that $\omega$  increases linearly in $\epsilon$. To confirm this, we again simulate the system for $T_x\neq T_y$ with small $\epsilon$ and it matches well with the theory (see Fig.[\ref{OverdampedOmega}]). This confirmation is indeed expected as the theory is developed for small $\epsilon$.  

However, at $\epsilon\rightarrow1$, according to Eq.[\ref{omega1}], $\omega$ decreases and when $\epsilon$ reaches to unity, $\omega$ becomes zero. Clearly, $\omega$ behaves non-monotonically when $\epsilon$ varies from $0$ to $1$.  This is confirmed by the simulation, as shown in Fig.[\ref{OverdampedOmega}]. This result seems counter-intuitive. Because in this case, the system is simultaneously connected to two heat baths at different temperatures. The degrees of freedom representing the system are also coupled with each other. So, there can be a heat flux driving the system out of equilibrium. However, $\omega$ still becomes zero and gyration is absent.  

By looking at the equation of motion given in Eq.[\ref{eom4}], one may get a clue to explain this seemingly counter-intuitive inference. By introducing an effective potential, $U_{\text{eff}}=\frac{1}{2}k(x^2+y^2+2\epsilon xy)$, one may rewrite Eq.[\ref{eom4}] as: $\gamma\dot x=-\frac{\partial}{\partial x}U_{\text{eff}}+\sqrt{2\gamma T_x}\xi_x$, and, $\gamma\dot y=-\frac{\partial}{\partial y}U_{\text{eff}}+\sqrt{2\gamma T_y}\xi_y$. Here, we note that -- (i) the linear superposition of  $\xi_x(t)$ and $\xi_y(t)$ is also a zero-mean Gaussian random variable, and, (ii) when $\epsilon=1$, $U_{\text{eff}}=\frac{1}{2}k(x+y)^2$. Hence, with $\epsilon=1$, after a long time, $\langle x+y \rangle$ relaxes to zero, which is the minima of the potential, $U_{\text{eff}}$. This implies that, for the stationary state, $\langle x\rangle=-\langle y\rangle$, and thus, the centre-of-mass of the particle is only fluctuating about the straight line $y=-x$, and it is not undergoing any gyration. This leads to $\omega=0$, when $\epsilon=1$ and $T_x\neq T_y$.  

Considering the non-monotonic behaviour of $\omega$, one can extremize it with respect to $\epsilon$ to obtain, $\epsilon^*\simeq \frac{1}{\sqrt{2}}=0.7071$, where the maximum value of $\omega$ is given by $\omega_{\text{max}}=\omega(\epsilon^*)$. This has also been confirmed by simulation. 

Hence, we conclude that an isotropically trapped, over-damped Brownian ellipsoid in 2D, subjected to two different temperatures $T_x$ and $T_y$ along $x$ and $y$ directions, exhibits gyration with a non-zero average angular velocity $\omega$. The Brownian gyration obtained here occurs due to an innate geometric as well as orientational anisotropy of the ellipsoid (represented by the parameter $\epsilon$) and a thermal asymmetry introduced via the two different temperatures, $T_x$ and $T_y$ of  two different thermal reservoirs. The system is connected with the reservoirs simultaneously. In the earlier works \cite{filliger2007brownian, dotsenko2013two, mancois2018two}, the gyration of an isotropic Brownian particle (Brownian sphere) was observed due to an externally applied anisotropic trapping potential. Whereas, here, the trapping potential is isotropic and the required anisotropy has been introduced by the intrinsic properties of the particle itself. In the next halve of this paper, we will discuss the case of gyration for a granular ellipsoid, where inertia of the ellipsoid will play a major role. 

\section{Gyration of a Granular Ellipsoid in 2D}

\subsection{The Model}

Granular materials are collections of discrete macroscopic solid grains with large enough sizes such that their Brownian motion due to thermal collisions with the particles of the surrounding medium becomes irrelevant. Hence, the dissipation in such systems is not related to the thermodynamic temperature, and the inertial effects become significant in the dynamics of such particles. It can be driven by external forces, which in this case will be stochastic. 

We consider a single granular ellipsoid constrained to move in 2D under the influence of an in-plane fluctuating force.  It is driven only by the fluctuating force, and no other forces (e.g. a trapping force, as discussed in the previous case) are applied externally on the particle. Such a system is motivated by the experiments with transverse motion of granular particles on a plate, subjected to vertical vibrations \cite{yamada2003coherent}. Our aim is to show analytically that the anisotropic granular particle can also exhibit gyration, if there exists a finite difference between the strengths of the two orthogonal fluctuations developed in the plane where the particle resides. 

The under-damped Langevin equations (for unit mass) governing the translational degrees of freedom of the granular ellipsoid are :

\begin{eqnarray}
\partial_t v_i=-\sum\limits_{j}\gamma_{ij}(\phi) v_j+f_i(t)
\label{eom5}
\end{eqnarray}

where, $v_i$ [with, $(i,j)\in(1,2)$] denotes the two components of the translational velocity of the centre-of-mass of the ellipsoid in the $x-y$ plane, i.e. $v_1=v_x, v_2=v_y$. Here, $\gamma_{ij}$ is the translational friction tensor (inverse of it being the translational mobility tensor, $\Gamma_{ij}$, defined earlier in section \ref{model}). The elements of the  friction tensor in terms of the longitudinal and transverse friction co-coefficients,  $\gamma_{\parallel}$ and $\gamma_{\perp}$ respectively, are given by \cite{dhont1996introduction}:
\begin{eqnarray}
\nonumber
\gamma_{xx}=\gamma_{\parallel}\cos^2\phi+\gamma_{\perp}\sin^2\phi\\
\nonumber
\gamma_{yy}=\gamma_{\parallel}\sin^2\phi+\gamma_{\perp}\cos^2\phi\\
\gamma_{xy}=\gamma_{yx}=\Delta\gamma \sin\phi\cos\phi
\end{eqnarray}
where, $\Delta\gamma=\gamma_{\perp}-\gamma_{\parallel}$.  Clearly, for a spherical particle, $\Delta\gamma = 0$, and this tensor becomes diagonal. For the ellipsoid, however, $\Delta\gamma$ is positive as $\gamma_{\parallel}<\gamma_{\perp}$. 

The externally applied Gaussian white noises are given by:

\begin{eqnarray}
\nonumber
f_x(t)=\sqrt{A_x}\xi_x(t)\\
f_y(t)=\sqrt{A_y}\xi_y(t)
\end{eqnarray}

where, $A_{x,y}$ are the translational noise strengths and $\xi_{x,y}$ are the zero-mean, Gaussian, white noises. These noise strengths are independent of the friction and thermodynamic temperature of the system as there exists no fluctuation-dissipation relation for granular systems. However, like the previous case, these athermal, externally applied fluctuations can retain similar statistical features of having the mean to be zero and their second moments being given by:

\begin{eqnarray}
&&\langle f_x(t)f_x(t') \rangle=A_x\delta(t-t')\\ 
&& \langle f_y(t)f_y(t') \rangle=A_y\delta(t-t')\\ \nonumber
\end{eqnarray}

The orientational dynamics of the ellipsoid can also be assumed to be under-damped:

\begin{eqnarray}
I\partial^2_t \phi+\gamma_r {\partial_t\phi}=\mathcal{M}
\label{eom6}
\end{eqnarray}
Here, $\gamma_r$ is the orientational friction co-efficient, and $I$ is the moment of inertia of the ellipsoid (about the axis perpendicular to $x-y$ plane, and going through the centre-of-mass of the ellipsoid). 

Similar to the colloidal particle, the total torque can be written as: $\mathcal{M}=\Lambda + f_r(t)=\Lambda + \sqrt{A_r}\xi_r(t)$. Here, $A_r$ is the strength of the fluctuations along the  orientational degree of freedom, and $\xi_r$ is the zero-mean Gaussian, white noise. Therefore, here also, $\langle f_r(t)\rangle=0$, and  $\langle f_r(t)f_r(t') \rangle=A_r\delta(t-t')$. However, the noise strength $A_r$ is also independent of $\gamma_r$ and the temperature of the system.   

\subsubsection{Small fluctuation in $\phi$ }

Due to a small but non-zero value of $\Lambda$, the probability distribution of $\phi$ and the corresponding momentum $p_{\phi}$, that is $P(p_{\phi},\phi)$, can be slightly asymmetric about $\phi=0$, which leads to a non-zero mean orientation of the ellipsoid, i.e. $\langle\phi\rangle \equiv \delta\phi\neq 0$. Here, $\delta\phi$ is small enough to keep terms $\mathcal{O}(\delta\phi)$ and neglect the terms $\mathcal{O}(\delta\phi^2)$ and all other terms of higher order in  Eq.[\ref{eom5}]. Hence, the $\phi-$averaged equation of motion becomes linear:

\begin{eqnarray}
\nonumber
\dot{v_x}=-\gamma_{\parallel}v_x-\epsilon v_y+\sqrt{A_x}\xi_x(t)\\
\dot{v_y}=-\gamma_{\perp}v_y-\epsilon v_x+\sqrt{A_y}\xi_y(t)
\label{eom7}
\end{eqnarray}

where, 

\begin{eqnarray}
 \epsilon=(\Delta\gamma)\delta\phi   
\end{eqnarray}

is the coupling constant.

\subsubsection{Small $\Delta\gamma$ limit}

Similar to the previous case, here also we consider $\Delta\gamma\to0$, for $\gamma_{\parallel}\approx\gamma_{\perp}$. Hence, $\epsilon<<1$, and both the longitudinal and transverse friction coefficients can be replaced by their average, denoted by $\gamma$.  Therefore the Eq.[{\ref{eom7}}] becomes:

\begin{eqnarray}
\nonumber
\dot{v_x}=-\gamma v_x-\epsilon v_y+\sqrt{A_x}\xi_x(t)\\
\dot{v_y}=-\gamma v_y-\epsilon v_x+\sqrt{A_y}\xi_y(t)
\label{eom8}
\end{eqnarray}

It is evident that $v_x$ and $v_y$ are coupled via a dissipative coupling $\epsilon$ as before. Also, anisotropies are introduced in the system via the shape and orientation of the particle and the difference between the fluctuation strengths, $A_x$ and $A_y$. In the current context, the main difference between the over-damped colloidal system and the under-damped granular system is that, there is no externally applied trapping force acting on the granular ellipsoid. However, the trap was necessary for the colloidal ellipsoid to gyrate.  

The mathematical forms of Eq.[\ref{eom8}] and Eq.[\ref{eom4}] are similar. The role of the dynamical variables $x$ and $y$ in Eq.[\ref{eom4}] are now played by $v_x$ and $v_y$ in Eq.[\ref{eom8}]. The role of the trapping strength $k$  in Eq.[\ref{eom4}] is now played by $\gamma$. However, unlike the strength of the harmonic trap $k$, the average friction coefficient $\gamma$ is an intrinsic property of the system itself. Finally, the role of  the noise strengths, $2T_x/\gamma$ and $2T_y/\gamma$, in the dynamics of the colloidal ellipsoid are now played by the noise strengths, $A_x$ and $A_y$. The coupling constants in both cases depend on the mean orientation of the ellipsoid and the difference between its longitudinal and transverse friction (or, mobility) coefficients.  

Due to this similarity, we expect the granular ellipsoid to gyrate as well. However, unlike the previous case,  it will occur in the 2D velocity-space without any externally applied trapping force. In the section below, while calculating the average angular velocity and the corresponding probability current density, we will confirm this.

\subsection{Results}

\subsubsection{Covariance Matrix and Velocity Variances}

The dynamics of the probability distribution for the velocity of the centre-of-mass of the granular ellipsoid, $P(v_x,v_y,t)$, is given by:

\begin{eqnarray}
\frac{\partial P(v_x,v_y,t)}{\partial t}=-\bf{\nabla.J}
\label{probdynv}
\end{eqnarray}

where, the probability current density ${\bf{J}}$ is given by, $J_i=\sum\limits_{j}\left(A_{ij}v_j P-\frac{1}{2}B_{ij}\frac{\partial P}{\partial v_j}\right)$, and the divergence is in the velocity-space. Here, $(i,j)\in (x,y)$ and the coefficient matrices $A_{ij}$ and $B_{ij}$ are given by,

\begin{eqnarray}
A &=&-\begin{bmatrix} \gamma & \epsilon \\ \epsilon & \gamma \end{bmatrix}\\
B &=&\begin{bmatrix} A_x & 0 \\ 0 & A_y \end{bmatrix}
\end{eqnarray}

Following a similar method as used in the colloidal case, we have the time-evolution of the covariance matrix, $\Sigma_g=\begin{bmatrix} \sigma_{11} & \sigma_{12} \\ \sigma_{21} & \sigma_{22}\end{bmatrix}$, from Eq.[\ref{probdynv}] as:  $\frac{d\Sigma_g}{dt}=A\Sigma_g+\Sigma_g A^T+B$.  Here, $\sigma_{11}=\langle v_x^2 \rangle$, $\sigma_{12}=\langle v_xv_y\rangle$, $\sigma_{21}=\langle v_yv_x \rangle$, and $\sigma_{22}=\langle v_y^2 \rangle$. In the stationary condition, we can obtain the velocity variances as:

\begin{eqnarray}
\nonumber
&&\sigma_{11} =\frac{2A_x+\epsilon_g^2 (A_y-A_x)}{4\gamma(1-\epsilon_g^2)}\\
\nonumber
&&\sigma_{22} =\frac{2A_y+\epsilon_g^2 (A_x-A_y)}{4\gamma(1-\epsilon_g^2)}\\
&&\sigma_{12}=\sigma_{21}=\frac{-\epsilon_g(A_x+A_y)}{4\gamma(1-\epsilon_g^2)}
\end{eqnarray}

where, $\epsilon_g\equiv\frac{\epsilon}{\gamma}(<< 1)$ is a dimensionless small parameter.  Small values of $\Delta\gamma$ and $\delta\phi$ ensure $\epsilon_g$ to remain much smaller than unity. As in the case of a colloidal particle, the higher order terms of $\epsilon_g$ are neglected in the expressions of the moments given above.

It is evident that the functional form of the velocity variances here are same as the functional form of the position variances in case of the colloidal particle, except the fact that the temperatures $T_x$ and $T_y$ in the colloidal case are now replaced by the arbitrary noise strengths $A_x/2$ and $A_y/2$ respectively in the granular case. In particular, 

\begin{eqnarray}
\nonumber
&&\sigma_{11}=\sigma_x\left(\frac{A_x}{2},\frac{A_y}{2},\gamma,\epsilon_g\right)\\
\nonumber
&&\sigma_{22}=\sigma_y\left(\frac{A_x}{2},\frac{A_y}{2},\gamma,\epsilon_g\right)\\
&&\sigma_{12}=\sigma_{xy}\left(\frac{A_x}{2},\frac{A_y}{2},\gamma,\epsilon_g\right)
\end{eqnarray}

and similarly for the ratio of $\sigma_{11}$ and $\sigma_{22}$, as shown below:

\begin{equation}
\mathcal{R}_g \equiv \frac{\sigma_{11}}{\sigma_{22}}=\frac{2-\epsilon_g^2(1-\frac{A_y}{A_x})}{2(\frac{A_y}{A_x})+\epsilon_g^2(1-\frac{A_y}{A_x})}={\mathcal{R}}\left(\frac{A_x}{2},\frac{A_y}{2},\epsilon_g\right)
\end{equation}

\begin{figure}[htp]
    \centering
    \includegraphics[width=8cm]{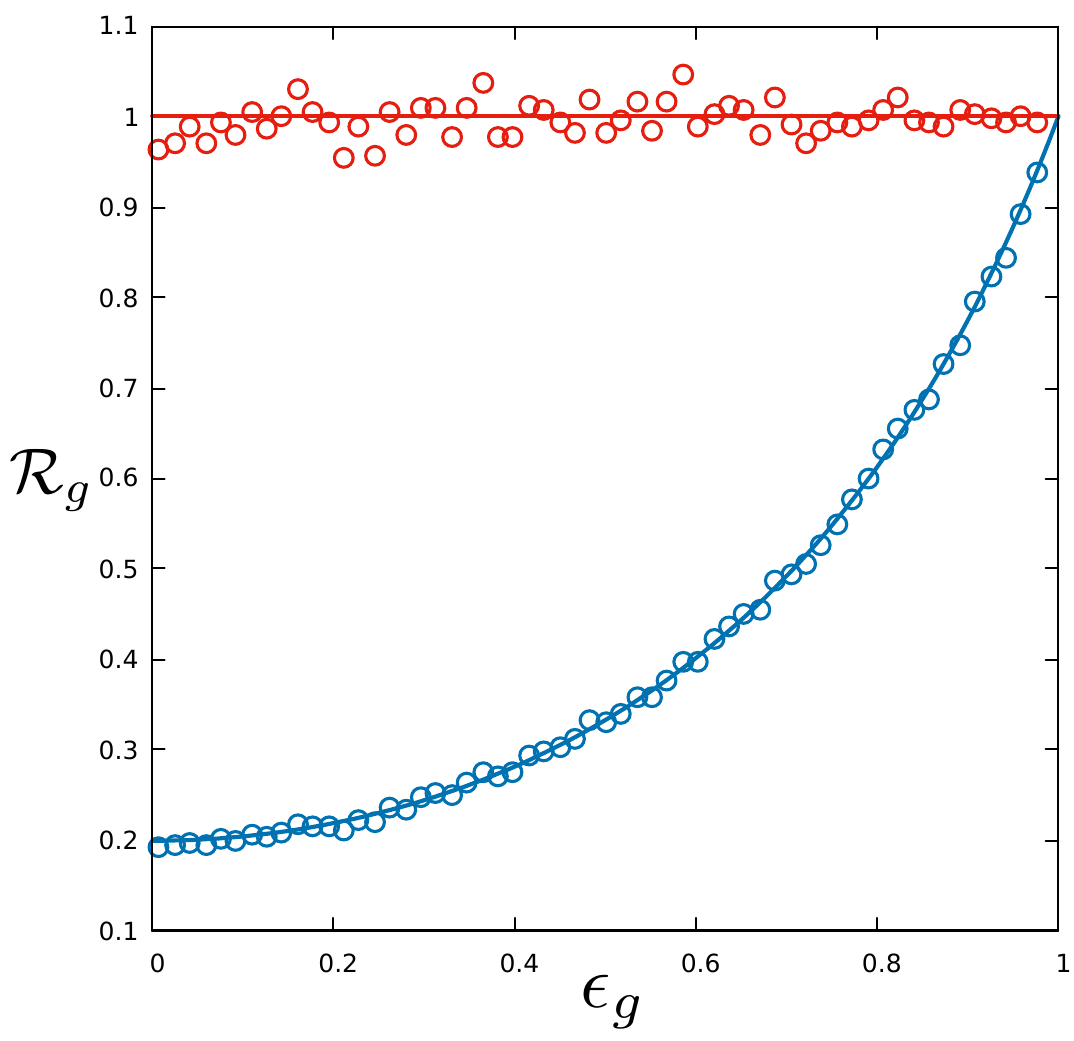}
    \caption{Variation of $\mathcal{R}_g$ with $\epsilon_g$. The red and blue circles are the numerical data obtained for $A_y=1$ and $A_y=5$ respectively (with $A_x=1$ for both the cases). The solid lines denote the analytical expressions.}
    \label{ratio_granular}
\end{figure}

Therefore, the characteristic properties and the limiting behaviours of the moments and their ratio $\mathcal{R}$ with respect to the coupling parameter $\epsilon$ in case of the colloidal ellipsoid are very similar to that of the granular ellipsoid with respect to $\epsilon_g$. We have simulated the dynamics of the granular ellipsoid and calculated $\mathcal{R}_g$.  In Fig.{\ref{ratio_granular}}, we have plotted $\mathcal{R}_g$ from both analytics and simulation and they match well.  

\subsubsection{Stationary Probability Distribution Function, $P(v_x,v_y)$}

Here, we will calculate the probability distribution function for the  translational velocity $(v_x,v_y)$  of the centre-of-mass of the granular ellipsoid, as we did in case of the Brownian gyrator in the previous section for the position of its centre-of-mass $(x,y)$. 

In the small $\epsilon_g$ limit, the dynamics of the velocity of the centre-of-mass of the ellipsoid is linear. The fluctuations involved in the dynamics are delta-correlated and Gaussian in nature. Hence, the stationary distribution can be taken to be a multi-variate Gaussian (which also satisfy the corresponding Fokker-Planck equation) given by:

\begin{equation}
P({\bf{v}})=\frac{e^{-\frac{1}{2}[({\bf{v}}-\langle{\bf{v}}\rangle)^T \Sigma^{-1}_g({\bf{v}}-\langle{\bf{v}}\rangle)]}}{2\pi\sqrt{{\text{Det}}\Sigma_g}}\equiv A_ge^{-p_g(v_x,v_y)}
\end{equation}

where, $\bf{v}=\begin{bmatrix} v_x\\v_y \end{bmatrix}$,  the normalisation constant is denoted by $A_g$ and the argument of the exponential is denoted by $p_g(v_x,v_y)$. Following the similar argument as before, the determinant and inverse of the covariance matrix $\Sigma_g$ can easily be obtained by replacing $(T_x,T_y,k,\epsilon)$ in ${\text{Det}}\Sigma$ and $\Sigma^{-1}$ by $\left(\frac{A_x}{2},\frac{A_y}{2},\gamma,\epsilon_g\right)$ respectively, i.e.  

\begin{eqnarray}
&& {\text{Det}}\Sigma_g={\text{Det}}\Sigma\left(\frac{A_x}{2},\frac{A_y}{2},\gamma,\epsilon_g\right)\\
&& \Sigma_g^{-1}=\Sigma^{-1}\left(\frac{A_x}{2},\frac{A_y}{2},\gamma,\epsilon_g\right)
\end{eqnarray}

where, terms beyond $\mathcal O(\epsilon_g^2)$ are discarded, as $\epsilon_g<<1$. Similarly, the normalization constant and the argument of the exponential can be obtained as:

\begin{eqnarray}
&& A_g = A\left(\frac{A_x}{2},\frac{A_y}{2},\gamma,\epsilon_g\right)\\
&& p_g(v_x,v_y)=\frac{1}{2}[({\bf{v}})^T \Sigma^{-1}_g({\bf{v}})]
\\ \nonumber &&
=\Delta_{1g} (v_x^2) + \Delta_{2g} (v_y^2) +\Delta_{3g} (v_xv_y)
\end{eqnarray}

where,
\begin{eqnarray}
\nonumber
\Delta_{1g}=\Delta_1\left(\frac{A_x}{2},\frac{A_y}{2},\gamma,\epsilon_g\right)\\
\nonumber
\Delta_{2g}=\Delta_2\left(\frac{A_x}{2},\frac{A_y}{2},\gamma,\epsilon_g\right)\\
\Delta_{3g}=\Delta_3\left(\frac{A_x}{2},\frac{A_y}{2},\gamma,\epsilon_g\right)
\end{eqnarray}

\begin{figure}[htp]
    \centering
    \includegraphics[width=8cm]{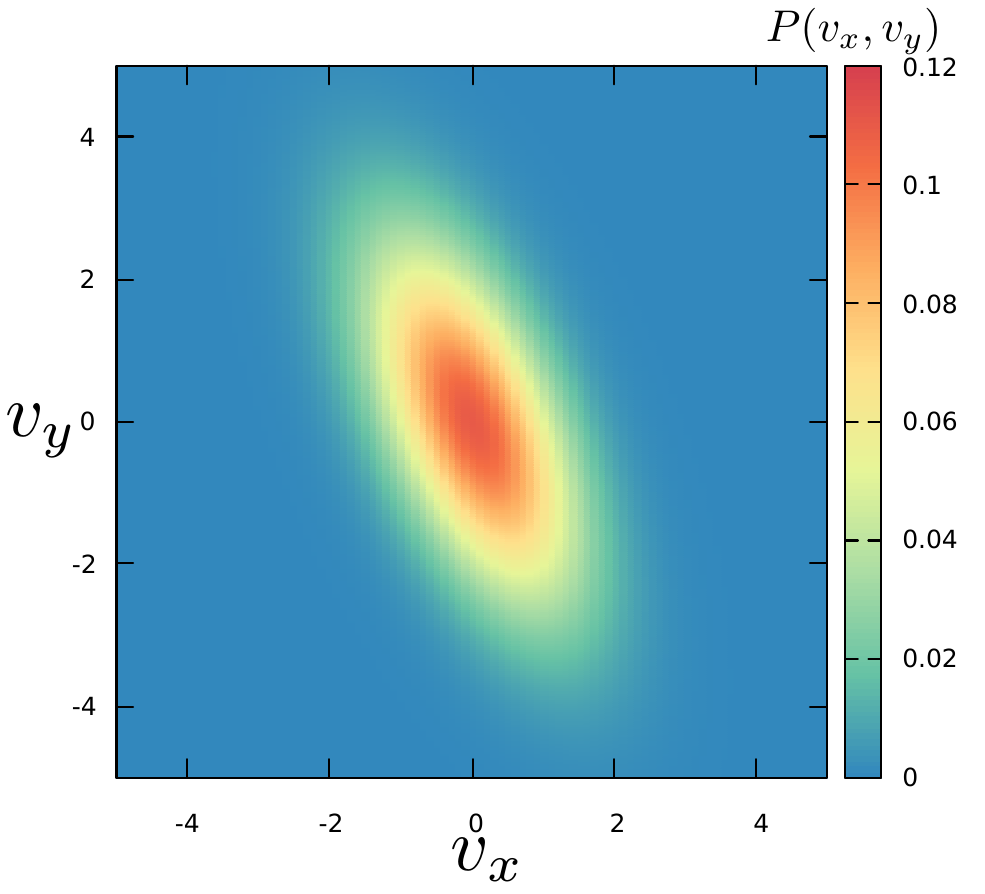}
    \caption{Contour plot of $P(v_x,v_y)$, for $\gamma=1, \epsilon_g=0.5, A_x=1, A_y=5$. Here, the colour scale denotes the various values of $P(v_x,v_y)$.}
    \label{probgran}
\end{figure}

Thus, we obtain the full stationary probability distribution $P({\bf{v}})$, as plotted in Fig.[\ref{probgran}].

In the polar representation, $v_x=v\cos\theta$, and $v_y=v\sin\theta$. Hence, the argument will be $p_g$ transformed as,

\begin{equation}
q_g(v,\theta)=v^2[\Delta_{g+}(1+\epsilon_g {\text{sin}}2\theta)+\Delta_{g-} {\text{cos}}2\theta]
\end{equation}

where,

\begin{eqnarray}
\Delta_{g+} = \frac{\Delta_{1g}+\Delta_{2g}}{2}\\
\Delta_{g-} = \frac{\Delta_{1g}-\Delta_{2g}}{2}
\end{eqnarray}

Similar to the previous case, by exponentiating $q_g$, we can obtain the probability density for the velocity of the center-of-mass of the granular ellipsoid in polar co-ordinates, $P(v,\theta)$, which will be used to evaluate the probability current densities in the next section.  

\subsubsection{Stationary Probability Current Density}

The components of the probability current densities along $v_x$ and $v_y$ can be represented in the polar co-ordinate system as:

\begin{eqnarray}
\nonumber
J_{v_x}(v,\theta)&=&-\gamma v\left( {\text{cos}}\theta+\epsilon_g {\text{sin}}\theta\right)P
\\ \nonumber &&
-\frac{A_x}{2}\left({\text{cos}}\theta\frac{\partial P}{\partial v}-\frac{{\text{sin}}\theta}{v}\frac{\partial P}{\partial \theta}\right)\\
\end{eqnarray}
\begin{eqnarray}
\nonumber
J_{v_y}(v,\theta)&=&-\gamma v({\text{sin}}\theta+\epsilon_g {\text{cos}}\theta)P\\
&-&\frac{A_y}{2}\left({\text{sin}}\theta\frac{\partial P}{\partial v}+\frac{{\text{cos}}\theta}{v}\frac{\partial P}{\partial \theta}\right)
\end{eqnarray}

Now, using $J_{v_\theta}(v,\theta)=J_{v_y}(v,\theta)\cos\theta-J_{v_x}(v,\theta)\sin\theta$, we get:

\begin{eqnarray}
\nonumber
J_{v_\theta} (v,\theta)&=&\frac{v\epsilon_g(A_y-A_x)}{2}[\Delta_{g+}(1+\epsilon_g {\text{sin}}2\theta)+\Delta_{g-} {\text{cos}}2\theta]\\ 
&\times & P(v,\theta)
\end{eqnarray}

Clearly $J_{v_{\theta}}$ is non-zero due to the asymmetry in fluctuations $(A_x\neq A_y)$ and the asymmetry in the intrinsic properties of the particle, such as shape $(\Delta\gamma\neq 0)$ and orientation $(\delta\phi\neq 0)$. This non-zero current density serves as an indicator of the non-equilibrium stationary state.

\subsubsection{Average Angular Velocity : Frequency of Gyration}

In polar co-ordinates, the ensemble average of the angular velocity  can be defined as: $\omega_g=\int_0^{2\pi} d\theta \int_0^{\infty}dv J_{v_\theta}(v,\theta)$, and therefore, we can obtain the ensemble-averaged frequency of gyration as (neglecting $\epsilon^3_g$ and terms of higher order):

\begin{equation}
\omega_g=\gamma\epsilon_g (A_y-A_x) \sqrt{\frac{1-\epsilon_g^2}{4A_xA_y+\epsilon^2_g(A_y-A_x)^2}}
\label{omegaG}
\end{equation}
If one neglects the terms of order $\epsilon_g^2$ as well, the above expression takes an even simpler form: $\omega=\frac{\gamma\epsilon_g}{2}(\sqrt{\frac{A_y}{A_x}}-\sqrt{\frac{A_x}{A_y}})$.

We note here that Eq.[\ref{omegaG}]  and Eq.[\ref{omega1}] have a similar mathematical form. Therefore, the characteristic features of $\omega$ and $\omega_g$ are similar too.  From Eq.[\ref{omegaG}], it is clear that the granular ellipsoid can gyrate if $A_x\neq A_y$ and the dissipative coupling $\epsilon_g$  remains between $0$ and $1$. However, if $A_x=A_y$, both the gyration current $J_{v_{\theta}}$ and the frequency of gyration $\omega_g$ will become zero for all $\epsilon_g$  ($0\leq \epsilon_g \leq 1$). Though the system will not be in the conventional, thermodynamic equilibrium as there exists no FDR. When $A_x\neq A_y$ but $\epsilon_g=0$, the translational degrees of freedom of the system become decoupled from each other. Hence, $J_{v_{\theta}}$ and $\omega_g$ becomes zero again, and there is no gyration. When $A_x\neq A_y$ but $\epsilon_g=1$, then also $J_{v_\theta}$ and $\omega_g$ both are zero for the similar reason as discussed in the previous case of the colloidal ellipsoid, when $T_x\neq T_y$ and $\epsilon=1$. When $A_x\neq A_y$, $\omega_g$ behaves non-monotonically with varying $\epsilon_g$ from $0$ to $1$. Also, the sense of the gyration (clockwise or anti-clockwise) gets flipped when $A_x$ becomes greater than $A_y$. 

As $\omega_g$ is non-monotonic between $\epsilon_g=0$ and $\epsilon_g=1$ it can be extremize it with respect to $\epsilon_g$ and obtain, $\epsilon_g^*\simeq \frac{1}{\sqrt{2}}=0.7071$, where $\omega_g$ becomes maximum ($=\omega_g(\epsilon^*)$). One may note here that the same value of the coupling constant was obtained in the colloidal case, where the average angular velocity attained its maximum,

Here, we have theoretically shown that when a granular particle is constrained to move in 2D and subjected to only zero-mean, Gaussian, white fluctuating forces, it generates directed motion, namely gyration, even without being trapped. The gyration with a non-zero average angular velocity and current leads towards the emergence of a rotational activity in the system. The gyration will occur if there exists : (i) an asymmetry in the fluctuating force, i.e. the amplitudes of the noisy forces along the two translational degrees of freedom of the particle must be different from each other, (ii) the mean orientation of the particle must be non-zero, and (iii) the friction between the particle and its surrounding medium must be asymmetric, i.e. it must be different along the different symmetry axis of the particle. The last two conditions are the intrinsic properties of a chiral, ellipsoidal, granular particle.

\begin{figure}[htp]
    \centering
    \includegraphics[width=8cm]{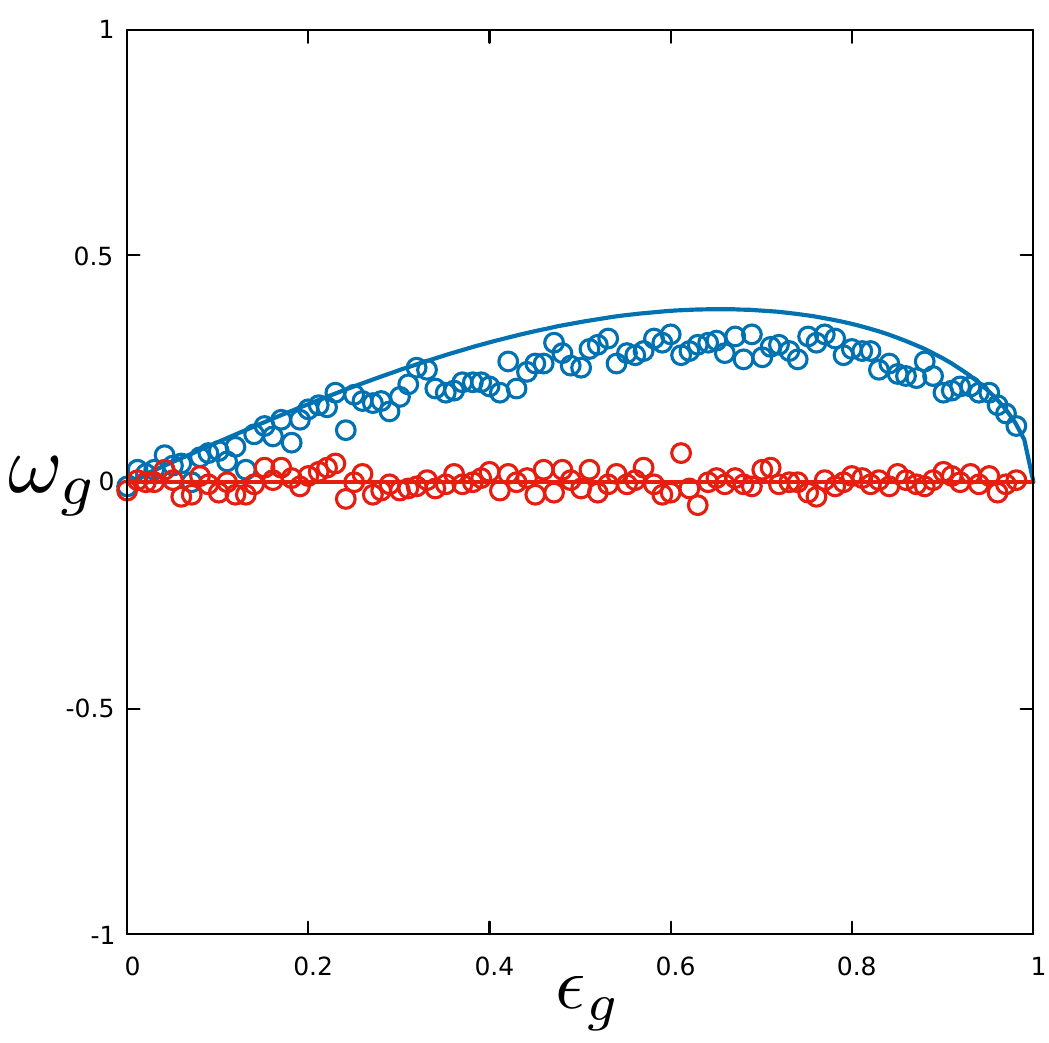}
    \caption{Variation of $\omega_g$ with $\epsilon_g$. The red and blue circles are the numerical data obtained for $A_y=1$ and $A_y=5$ respectively (with $A_x=1$, $\gamma=1$ for both the cases). The solid lines denote the analytical expressions.}
    \label{omegaGran}
\end{figure}

\subsubsection{Average Angular Velocity: The Role of Pseudo-Forces}

In this section, we will discuss an aspect which is exclusive for systems where the inertia is significant. By now, it is theoretically evident that a granular ellipsoid with an asymmetry in the shape (and a non-zero mean orientation), gyrates when subjected to asymmetric fluctuations. The average frequency of gyration $\omega_g$ is calculated and given in Eq.[\ref{omegaG}]. Due to this gyration, the particle will experience pseudo-forces, calculated from the inertial (laboratory) frame. The three major pseudo-forces are the: (i) Euler force, (ii) centrifugal force, and (iii) Coriolis force. In the stationary state, $\omega_g$ is independent of time, and hence, Euler force will be absent. The centrifugal force goes as $\omega_g^2$, and hence, it will be $\mathcal{O}(\epsilon_g^2)$.  However, the Coriolis force goes as $\omega_g$, and therefore, it is $\mathcal{O}(\epsilon_g)$. Consequently, the Coriolis force in the system dominates over the centrifugal force and therefore it can be neglected for the system under consideration. If the average translational velocity of the centre-of-mass of the ellipsoid is ${\bf{v_g}}$, then from the definition of the Coriolis force : $\frac{d{\bf{ v}}_g}{dt} \simeq 2{\bf{\omega}}_g {\bf{v}}_g$ (for unit mass). Using this, we directly compute $\omega_g$ by simulating the system with different $\epsilon_g$. This result is plotted in Fig.[\ref{omegaGran}] with the analytically obtained $\omega_g$ in Eq.[\ref{omegaG}], by varying $\epsilon_g$ from $0$ to $1$. The numerically and analytically obtained $\omega_g$ agree well. Hence, it has become evident that the granular ellipsoid gyrates mostly due to the Coriolis force generated in the system. Therefore, the inertial system that we are concerned about is a table-top set-up for generating Coriolis force from asymmetric fluctuations.

\section{Second-order correction in the orientational fluctuation}

So far we considered that the orientational fluctuations for both overdamped and underdamped cases are so small that considering the terms linear in $\delta\phi$ was enough. Now, one can also consider the fluctuations to be slightly higher in magnitude, such that one has to retain the next higher-order term after $\delta\phi$, that is, $(\delta\phi)^2$.  Then, for the overdamped case, the dynamics is modified  as: $\frac{dx}{dt} = -k[(\Gamma_\parallel-\Delta\Gamma(\delta\phi)^2)x+(\Delta\Gamma\delta\phi)y]+\sqrt{2k_B T_x(\Gamma_\parallel-\Delta\Gamma(\delta\phi)^2)}\xi_x(t)$, and, $\frac{dy}{dt} = -k[(\Gamma_\perp+\Delta\Gamma(\delta\phi)^2)y+(\Delta\Gamma\delta\phi)x]+\sqrt{2k_B T_y(\Gamma_\perp+\Delta\Gamma(\delta\phi)^2)}\xi_y(t)$. One can readily observe that the previous form of the dissipative coupling is retained even after including the second-order term, $(\delta\phi)^2\equiv \langle\phi^2\rangle - \langle\phi\rangle^2$. However, the remaining co-efficients have been modified by a small amount, as $\delta\phi$ itself is very small. Hence, the overall quantitative nature of gyration will not be affected much in the regime of higher orientational fluctuations. Interestingly, as $(\delta\phi)^2\sim T_r$ (the temperature associated with the rotational dynamics), the $\phi-$averaged dynamics of the centre-of-mass of the ellipsoid will now include the effect of orientational fluctuations also, which can be different from $T_x$ and $T_y$, in general.   

Similarly, the translational Langevin equations in the underdamped case will be modified due to the second-order orientational fluctuations as, $\frac{dv_x}{dt} = -[(\gamma_\parallel+\Delta\gamma(\delta\phi)^2)v_x+(\Delta\gamma\delta\phi)v_y]+\sqrt{A_x}\xi_x(t)$, and, $\frac{dv_y}{dt} = -[(\gamma_\perp-\Delta\gamma(\delta\phi)^2)v_y+(\Delta\gamma\delta\phi)v_x]+\sqrt{A_y}\xi_y(t)$.
Here also, the coupling between $v_x$ and $v_y$ are retained as before. Hence, no change is expected in the quantitative behaviour of the gyration in the granular scale. Also, the strength of the orientational fluctuations $A_r$ will now affect the dynamics of the centre-of-mass of the granular ellipsoid via $(\delta\phi)^2$.   

\section{A Note on the Numerical Method}

The coupled Langevin equations for translational motion (in both overdamped and underdamped cases) have been solved numerically using the finite-difference scheme with a time step-size of $10^{-3}$. Doubly-precise values of the relevant quantities have been obtained for $10^{4}$ different values of the coupling parameter $\epsilon$, ranging between $0$ to $1$. For each value of $\epsilon$, the quantity under consideration has been averaged over $10^{7}$ realisations. For the white noises, random numbers have been picked up from a Gaussian distribution having zero mean and unit variance.

\section{Concluding Remarks}

To conclude, we will summarise the work here. First, we have considered a Brownian ellipsoid in 2D with a finite tilt in its orientation. It is trapped in an isotropic harmonic potential and subjected to two distinct temperatures along each of the space dimensions. A dissipative coupling between the space co-ordinates arises due to the difference between the longitudinal and transverse mobility and a finite tilt in the orientation of the ellipsoid. We have shown that such a Brownian ellipsoid can gyrate. After calculating the position probability distribution function, we have calculated the average frequency of gyration analytically in the small coupling limit. The results have also been verified numerically.  From this part we learn that anisotropic fluctuations, which is the fundamental requirement for BG, can be produced by the particle itself (and not by the trap) by using the inherent asymmetry in its shape and orientation. 

In the other halve of the paper, we consider an inertial granular ellipsoid in 2D with a finite tilt in its orientation. Unlike the previous case, the particle is free to move in 2D, i.e. there is no trapping potential acting on the particle. However, the particle is subjected to externally applied athermal (but zero-mean Gaussian and white) fluctuations with different strengths along the different space dimensions. Here we have shown that due to the difference between the longitudinal and transverse friction together with the finite tilt in the orientation of the granular ellipsoid, the velocity components of the centre-of-mass of the ellipsoid along the space dimensions  get coupled. This dissipative coupling and the difference between the fluctuation strengths together induces gyration in the dynamics of the ellipsoid. We have calculated the velocity probability distribution function and the average frequency of gyration of the ellipsoid analytically. We have also verified the results numerically. Interestingly, we have shown that the Coriolis force plays a major role in producing gyration at the granular scale. Here we learn that to generate gyration in the velocity space, an inertial and orientationally biased granular ellipsoid does not even require any trap.

\begin{figure}[htp]
    \centering
    \includegraphics[width=7.5cm]{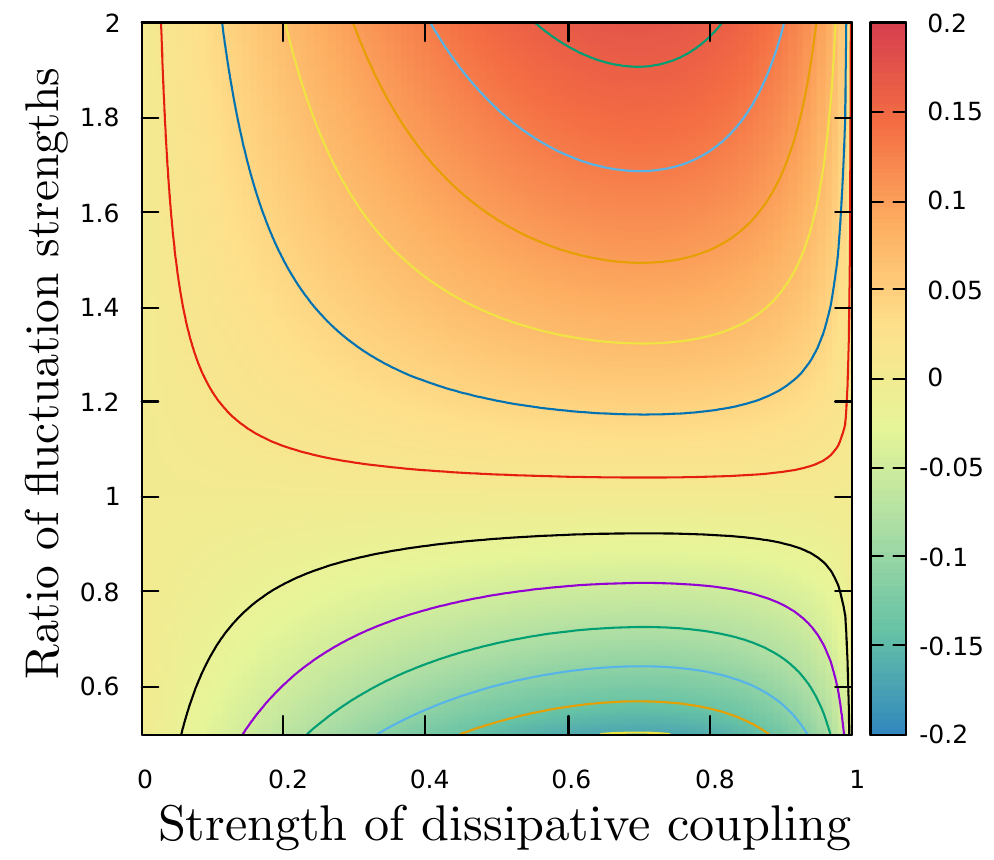}
    \caption{Heat-map of the average angular velocity (shown in the colour scale) with the ratio of the fluctuation strengths ($\frac{T_y}{T_x}$ or, $\frac{A_y}{A_x}$) and the strength of the dissipative coupling ($\epsilon$ or, $\epsilon_g$). The regions where the value of the ratio is greater or lesser than unity are clearly depicted. The solid lines of various colours denote the iso-angular velocity contours.}
    \label{heatmap}
\end{figure}

In Fig.[\ref{heatmap}], we consider the plane of the dimensionless coupling constant and the ratio of the strengths of the fluctuations along the space dimensions. In this plot, we have shown the regions where the ellipsoid (both over-damped as well as under-damped) can gyrate and where it cannot, by simply plotting the heat-map of the average gyration frequency. This plot gives us a complete picture of the factors involved in the emergence of gyration, be it in the colloidal or in the granular scale.

After summarising the findings of this paper, one may find it important to think about a few points regarding our study. First, taking the advantage of the gyration of the ellipsoid, one may think of extracting thermodynamic work from the system. For example, one may attach a tiny crankshaft to convert the gyration into a linear motion and then attach a tiny piston with the shaft such that it can do work on a given load. However, the fluctuations about the mean gyration frequency should significantly affect the efficiency of such a stochastic machine. Recently, the fluctuations about the mean gyration frequency of a spherical Brownian particle in an anisotropic trap have been analysed in detail \cite{viot2023destructive}. The fluctuations seem to be quite large in this case. This will affect the work output of such a system. Improvements towards a more regular gyration are also suggested in \cite{viot2023destructive}.

\begin{figure}[htp]
    \centering
    \includegraphics[width=9.0cm]{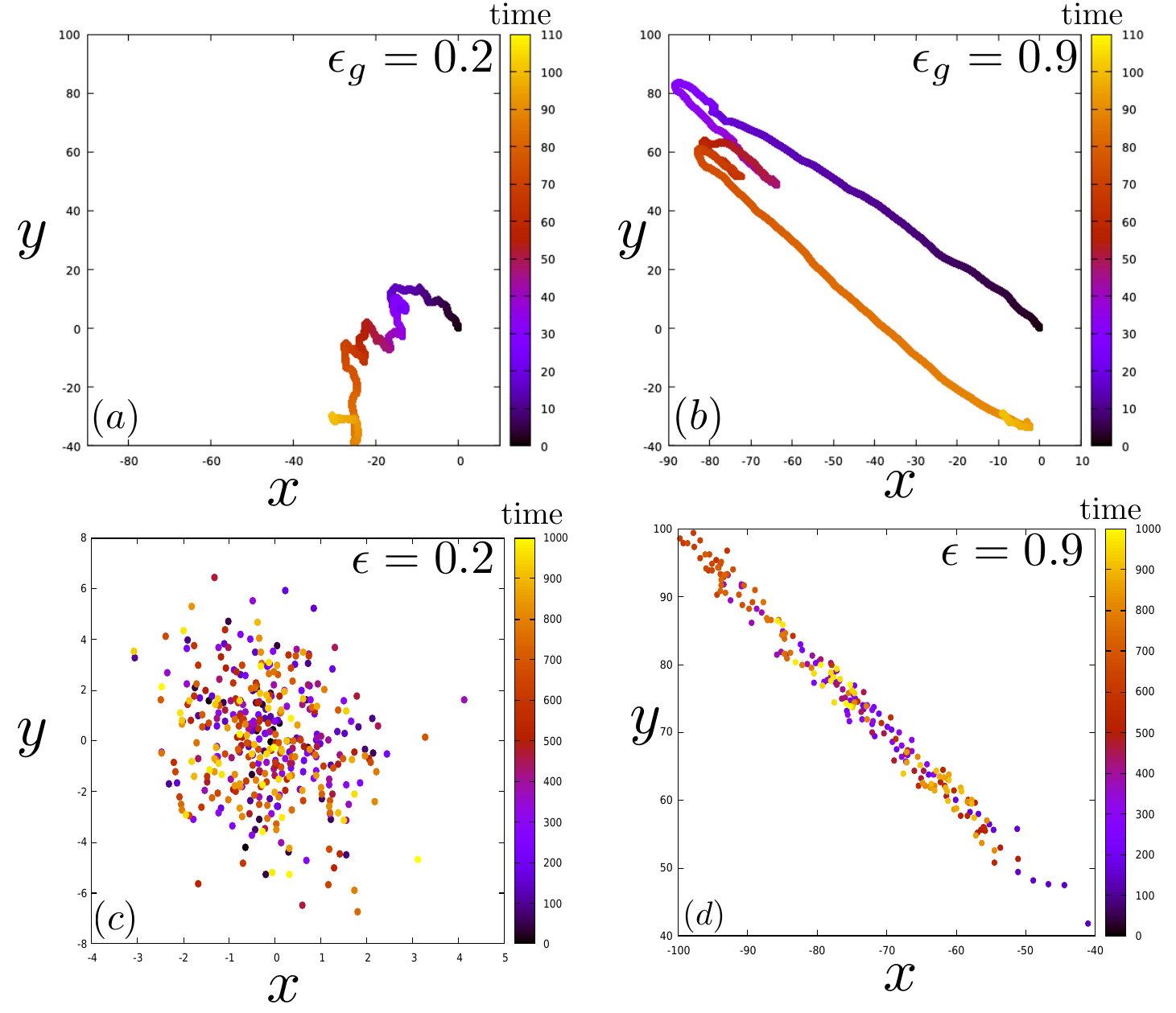}
    \caption{Trajectories of the centre-of-mass of the ellipsoid. (a,b) : Granular, inertial ellipsoid on the $xy$ plane for two different values of the coupling constant, $\epsilon_g = 0.2, 0.9$ respectively. (c,d) :  Brownian, over-damped ellipsoid on the $xy$ plane for two different values of the coupling constant, $\epsilon = 0.2, 0.9$ respectively. Time is denoted by the colour-scale.}
    \label{traj}
\end{figure}

In Fig.[\ref{traj}], two typical trajectories of the centre-of-mass of the gyrating granular ellipsoid for two different values of $\epsilon_g$ are plotted in the upper panel. Similarly, in the lower panel, two typical trajectories of the centre-of-mass of the gyrating Brownian ellipsoid for two different values of $\epsilon$ are plotted. The fluctuations of the trajectories of the granular ellipsoid seem to be much smaller than that of the Brownian ellipsoid. This suggests that the fluctuations about the mean gyration frequency $\omega_g$ in the granular case should also be smaller. This can facilitate the granular system to have a higher work output, and consequently, a higher efficiency.       

Secondly, one may also note that here we have assumed $\gamma_{\perp}\simeq\gamma_{\parallel}$, for which the individual friction coefficients are replaced by their average. In general, this is not true. For example, in case of rods, $\gamma_{\perp}>\gamma_{\parallel}$ and one cannot replace them by their average. Hence, in general, $x$ and $y$ are coupled  with different coupling constants $c_1$ and $c_2$ (see Eq.[\ref{eom3}]) in case of a Brownian ellipsoid. Similarly, in case of a granular ellipsoid, the the two different dimensionless coupling constants will be $\epsilon_{g_1}=\frac{\epsilon}{\gamma_{\parallel}}$ and $\epsilon_{g_2}=\frac{\epsilon}{\gamma_{\perp}}$. Optimising the fluctuations about the mean gyration frequency with these parameters will be an important aspect to explore the characteristics of the work output and efficiency of such a stochastic machine. 

Third, one may also think of an electrical analogue of the colloidal (granular gyrator) respectively based on the capacitor-resistor (inductance-resistor) circuits with two different sources of temperature-dependent electrical fluctuations. 

Finally, based on the concept of microscopic gyration, one may build up {\it{artificially intelligent}} Braitenberg-like machines which can control its dynamics by varying its own shape. In particular, one can think of an ellipsoid that can gyrate by {\it{sensing}} the difference between the strengths of fluctuations ($\equiv D$, say) along the space dimensions (i.e., $D=|A_x-A_y|$ or, $|T_x-T_y|$). Suppose the sensor of the machine is connected to the coupling constant ($\epsilon_g$ in the granular case  or, $\epsilon$ in the colloidal case) in such a way that when $D$ is higher than a certain threshold value $D_c$, the coupling constant $\epsilon_g$ (or, $\epsilon$) will be set to a non-zero value between $0$ and $1$ (i.e., the particle becomes a spherically-anistropic ellipsoid), and otherwise it is set to zero (i.e., the particle remains spherically symmetric). In other words, the particle can transform its shape dynamically as well as reversibly from a sphere to an ellipsoid and vice-versa, depending on the value of $D$ it has sensed.  Hence, one can expect that the particle will exhibit gyration only when $D>D_c$, and otherwise there is no gyration and the particle will simply diffuse. Consequently, in the region where $D>D_c$, the particle will gyrate and being {\it{affectionate}} to the region, it will remain there for a longer span than the region where $D<D_c$.  This mechanism should assist the particle to navigate in a complex, inhomogeneous environment, particularly where the strength of the fluctuations along different degrees of freedom are different. One may note here that the essential ingredient  of such a navigation is the smooth and reversible shape-transformation of the particle. Many living cells are intrinsically capable of  controlling their shape by complex bio-chemical pathways. Responsive colloids are promising synthetic materials for such a shape-transformation  \cite{tu2014shape,lu2011thermosensitive,brijitta2019responsive,baul2021structure}, which can be utilised for navigation and targeted delivery. Research along these directions are in progress.

\section{Acknowledgements} 

Authors thank the start-up grant from University Grants Commission (UGC) via the UGC Faculty Recharge Program (UGCFRP) and the Core Research Grant (CRG/2019/001492) from SERB, India. AS thanks Abhishek Chaudhuri, Anweshika Pattanayak, and Sourabh Lahiri for their critical comments on the work. 

\bibliographystyle{unsrt}
\bibliography{sample.bib}
\end{document}